\documentclass[aps,prl,twocolumn,longbibliography,floatfix,10pt,superscriptaddress]{revtex4-2}
\usepackage[utf8]{inputenc}
\usepackage{mathrsfs}
\usepackage{natbib}
\usepackage{graphicx}
\usepackage{latexsym}
\usepackage{amssymb}
\usepackage{amsmath}
\usepackage{amsfonts}
\usepackage{gensymb}
\usepackage{bm}
\usepackage{hyperref}
\usepackage{braket}
\usepackage{physics}
\usepackage{bbold}
\usepackage[caption=false]{subfig}
\usepackage[dvipsnames]{xcolor}
\usepackage[normalem]{ulem}
\usepackage{siunitx}
\usepackage{mhchem}
\usepackage{multirow}

\usepackage{slashed}

\definecolor{myforestgreen}{RGB}{34,139,34}

\newcommand{\supplementarysection}{%
  \setcounter{figure}{0}
  \let\oldthefigure\thefigure
  \renewcommand{\thefigure}{S\oldthefigure}
  \setcounter{section}{0}
  \let\oldthesection\thesection
  \renewcommand{\thesection}{S\oldthesection}
  \setcounter{equation}{0}
  \let\oldtheequation\theequation
  \renewcommand{\theequation}{\oldtheequation}
  \setcounter{table}{0}
  \let\oldthetable\thetable
  \renewcommand{\thetable}{S\oldthetable}
}

\newcommand{\bi}{\begin{itemize}}
\newcommand{\ei}{\end{itemize}}
\newcommand{\be}{\begin{eqnarray}}
\newcommand{\ee}{\end{eqnarray}}
 
\newcommand{\Eq}[1]{Eq.~(\ref{#1})}
\newenvironment{dfn}{{\vspace*{1ex} \noindent \bf Definition }}{\vspace*{1ex}}

\newcommand{\nn}{\nonumber}  %

	\newcommand{\beq}{\begin{eqnarray}}
	\newcommand{\eeq}{\end{eqnarray}}
	\newcommand{\bea}{\begin{eqnarray}\begin{aligned}}
	\newcommand{\eea}{\end{aligned}\end{eqnarray}}

	\renewcommand{\v}{{\bf v}}

\begin{document}

\title{Interacting Chern insulator transition on the sphere: revealing the Gross--Neveu--Yukawa criticality}

\author{Zhi-Qiang Gao}
\thanks{These authors contributed equally.}
\affiliation{Department of Physics, University of California, Berkeley, CA 94720, USA \looseness=-2}
\affiliation{Material Science Division, Lawrence Berkeley National Laboratory, Berkeley, CA 94720, USA \looseness=-2}
\author{Taige Wang}
\thanks{These authors contributed equally.}
\affiliation{Department of Physics, University of California, Berkeley, CA 94720, USA \looseness=-2}
\author{Dung-Hai Lee}
\affiliation{Department of Physics, University of California, Berkeley, CA 94720, USA \looseness=-2}
\affiliation{Material Science Division, Lawrence Berkeley National Laboratory, Berkeley, CA 94720, USA \looseness=-2}


\begin{abstract}

In two dimensions a Chern insulator can be tuned from a topological state to an ordinary band insulator without breaking any symmetry, offering a clean example of ``beyond-Landau'' criticality. We show that electron interactions drive this transition to the Gross–Neveu–Yukawa (GNY) fixed point, a benchmark problem of broad interest across high-energy and condensed-matter physics. By solving the minimal Dirac model on the surface of a sphere, we obtain the full spectrum of critical excitations from systems containing only a few electrons. The leading critical exponents match state-of-the-art bootstrap bounds, and several previously unseen collective modes appear. Our results demonstrate that placing a Dirac cone on a curved geometry provides a surprisingly simple and accurate setup for strongly interacting topological transitions.

\end{abstract}

\maketitle

\begin{figure}[t]
  \centering
  \includegraphics[width=0.6\linewidth]{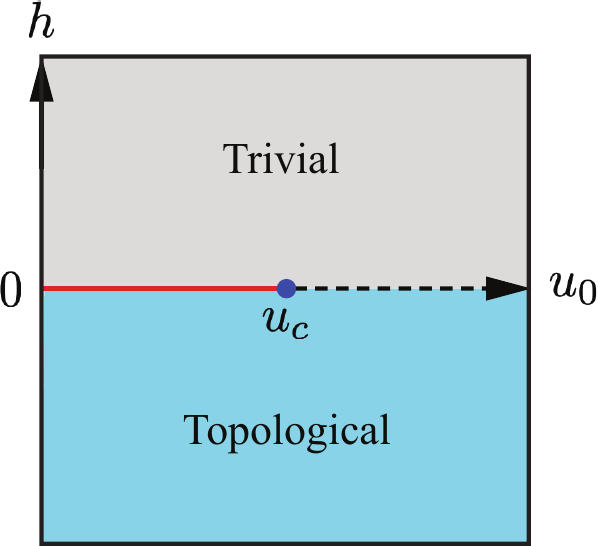}
  \caption{Mean-field phase diagram in the \((h,u_0)\) plane. 
  Topological \((h<0)\) and trivial \((h>0)\) phases meet via a continuous transition for \(u<u_c\) (red line), described by a free-fermion CFT, 
  which becomes first-order (dashed line) for \(u>u_c\). 
  The point \(u=u_c\) (blue dot) is a GNY transition where fermions and the mass operator both get gapless.}
  \label{fig1}
\end{figure}

\textbf{Introduction.}
Quantum phases can differ in topology even when they share all microscopic symmetries.  
Because such changes of topology fall outside the Landau framework based on symmetry breaking, they have become touchstones for modern theories of phase transitions.  
A widely discussed two–dimensional example is the passage from a trivial band insulator to a Chern insulator: as one tunes a control parameter, the bulk gap collapses while a single chiral edge channel and a quantised Hall response switch on~\cite{PhysRevB.74.235111,Wen01101995}.

The essential low-energy degrees of freedom are those of a single Dirac cone.  
Introducing the shorthand  
\(\boldsymbol{\sigma} \cdot \nabla = \sigma_x\partial_x + \sigma_y\partial_y\),
we collect kinetic, mass, and interactions into one Hamiltonian  
\begin{equation}
\tilde H = \int d^{2}x
\Bigl[
      \psi^{\dagger}\bigl( -i\,\boldsymbol{\sigma} \cdot \nabla + h \sigma_z \bigr)\psi
      - \frac{u_0}{2}\bigl( \psi^{\dagger}\sigma_z\psi \bigr)^{2}
\Bigr].
\label{eq:Ham}
\end{equation}
Changing the sign of the single-particle mass \(h\) changes the Chern number by \(\pm1\) and produces a free-fermion conformal field theory (CFT) at \(h = 0\).  
For weak interaction \(u_0\) the quartic term is irrelevant, so the transition remains in that free-Dirac class.  
At stronger coupling \(u_c\) the system develops a spontaneous mass  
\(\langle \psi^{\dagger}\sigma_z\psi \rangle \neq 0\); the point where this ordering collides with the band-inversion line is a multicritical fixed point that belongs to the Gross–Neveu–Yukawa (GNY) universality class, as sketched in Fig.~\ref{fig1}.

A Hubbard–Stratonovich decoupling of Eq.~\eqref{eq:Ham} maps the lattice model exactly onto the GNY action (see End Matter for details),
\begin{equation}
\mathcal{L}_{\text{GNY}} =
\bar\psi_\alpha \bigl(i\gamma^\mu\partial_\mu - g\phi\bigr)\psi_\alpha
+ \frac12(\partial_\mu\phi)^2
+ \frac{t}{2}\phi^2
+ \frac{\lambda}{4}\phi^4,
\end{equation}
where a single Dirac fermion $\psi$ is Yukawa-coupled to an Ising-like scalar field $\phi$. The GNY point is of considerable interest to both high-energy and condensed-matter communities, including chiral symmetry breaking in QCD~\cite{Gross1974}, Kekulé ordering in graphene~\cite{PhysRevB.80.075432}, and nodal superconductivity~\cite{PhysRevLett.85.4940}.  
Analytical and numerical studies—conformal bootstrap~\cite{Iliesiu2016,Iliesiu2018,Erramilli2023}, \(\epsilon\) expansions~\cite{PhysRevB.98.125109}, large-\(N\) techniques~\cite{PhysRevD.103.125004,Fei2016}, and sign-problem-free quantum Monte Carlo (QMC)~\cite{PhysRevLett.128.225701,PhysRevD.101.074501,PhysRevB.101.064308,PhysRevB.108.L121112}—agree on the leading critical exponents yet disagree on many sub-leading operators, leaving the full spectrum unsettled.

A faithful numerical treatment of an interacting \emph{single} Dirac cone is notoriously difficult for four distinct reasons. 
First, any local, time-reversal-symmetric lattice regularization suffers from the parity anomaly~\cite{NIELSEN1981173}. Standard approaches enlarge the flavor number~\cite{PhysRevD.101.074501,PhysRevB.101.064308}, break time-reversal symmetry~\cite{PhysRevD.10.2445}, or adopt non-local hoppings~\cite{PhysRevLett.128.225701}, each of which can obscure the intended universality class. 
Second, the critical interacting Dirac fermion is intrinsically strongly correlated. It cannot be reached perturbatively from the free Dirac fermion CFT, as the four-fermion interaction is irrelevant at the free fixed point. 
Third, without $SO(3)$ symmetry, most microscopic calculations can only extract critical exponents from order parameters and correlation functions instead of using state-operator correspondence to read out scaling dimensions of full operator content of the CFT~\cite{Zhu2023}. They also typically suffer from relatively large numerical error due to the lack of enough symmetry.
Finally, even on the sphere that preserves $SO(3)$, maintaining the linear dispersion of the Dirac fermion is nontrivial: standard spectrum-based fuzzy sphere constructions~\cite{Zhu2023,Hu2023OPE,Han2023FourPoint,Zhou2024SO5,
Han2024O3,Hu2024Defect,Fan2025,Zhou2026Fermionic} are not suitable for Dirac fermions, since the monopole at the sphere center generates Landau levels instead of the target linear spectrum.

Here we sidestep all these obstacles by formulating the interacting Dirac cone directly on the sphere, which generalizes the fuzzy sphere framework to a critical point of interacting Dirac fermion. Curvature supplies the necessary spin connection, while the overall geometry—with its full rotational symmetry—greatly reduces finite-size effects.  
Exact diagonalization of systems containing fewer than ten electrons already yields excitation spectra whose level spacings translate directly into operator dimensions.  
The leading values coincide with the best bootstrap bounds, and several higher-dimension primaries, including an almost-conserved charged current, appear for the first time.  
Spherical geometry therefore offers a compact and transparent numerical microscope for the Gross--Neveu--Yukawa fixed point and for Dirac criticality more broadly.

\begin{figure}[htb]
  \centering
  \includegraphics[width=\linewidth]{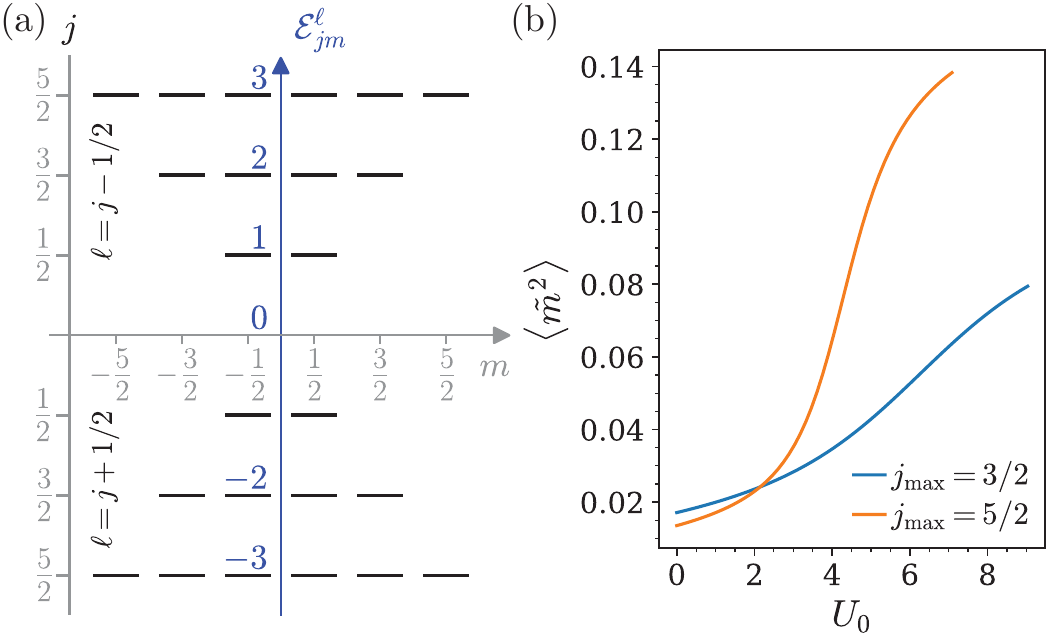}
  \caption{(a) Single-particle energy levels \(\mathcal{E}_{jm}^\ell\) of free Dirac fermions on the sphere. Levels up to \(j \le j_{\mathrm{max}} = 5/2\) are shown. (b) Rescaled order parameter \(\langle \tilde{m}^2\rangle\) of the ground state across the paramagnetic-to-ferromagnetic transition as a function of the contact interaction \(U_0\) (see main text for details).}
  \label{fig2}
\end{figure}

\textbf{Dirac fermions on $\mathbb{S}^2$.} Setting the radius of the sphere to 1, the Dirac operator takes the form
\begin{equation}
D = i\gamma^\mu\nabla_\mu,
\end{equation}
where \(\nabla_\mu = \partial_\mu + \Omega_\mu\) is the covariant derivative for spinor fields, and \(\Omega_\mu\) is the spin connection. On $\mathbb{S}^2$, this connection is constant, leading to~\cite{Gutierrez2018}
\beq
D = \boldsymbol{\sigma}\cdot\mathbf{L} + 1
  = \mathbf{J}^2 - \mathbf{L}^2 + \frac14,
\eeq
where  \(\mathbf{L}\) denotes the orbital angular momentum, and \(\mathbf{J} = \mathbf{L} + \tfrac12\boldsymbol{\sigma}\) is the total angular momentum. From this operator, we construct the free Dirac Hamiltonian
\begin{equation}
H_0 = \int d\Omega\,\psi^\dagger(\Omega)D(\Omega)\psi(\Omega)
      + h\psi^\dagger(\Omega)\sigma_z\psi(\Omega).
\end{equation}
Besides the spatial \(\mathrm{SO}(3)\) and global charge \(\mathrm{U(1)}\) symmetries, this Hamiltonian also features three discrete \(\mathbb{Z}_2\) symmetries. When \(h=0\), it possesses a time-reversal \(\mathcal{T}\) symmetry that spontaneously breaks across the transition. It also admits charge conjugation \(\mathcal{C}\) (though \(\mathcal{C}\) does not commute with the global \(\mathrm{U(1)}\)). A spacetime parity \(\mathcal{P}\) symmetry is also present, but it does not commute with \(\mathrm{SO}(3)\). We elaborate on how \(\mathcal{C}\), \(\mathcal{T}\), and \(\mathcal{P}\) act on the fermionic operators in the Supplementary Material~\cite{supp}. Crucially, preserving these symmetries at the GNY criticality ensures that only one relevant operator remains. In particular, maintaining \(\mathcal{T}\) in the interacting regime confines the system to the \(h=0\) line of Fig.~\ref{fig1}, enabling the multicritical point to be approached by tuning just a single parameter \(u_0\).

The single-particle eigenstates of \(D\) are the spinor spherical harmonics \(\mathcal{Y}_{jm}^\ell\), satisfying
\begin{equation}
D\,\mathcal{Y}_{jm}^\ell= \mathcal{E}_{jm}^\ell \mathcal{Y}_{jm}^\ell, \quad \mathcal{E}_{jm}^\ell = \pm\bigl(j+\frac12\bigr) \text{ for } \ell=j\mp\frac12,
\end{equation}
with \(m=-j,-j+1,\dots,j\) and \(j=\tfrac12,\tfrac32,\dots\). We show the single-particle spectrum $\mathcal{E}_{jm}$ in Fig.~\ref{fig2} (a). These solutions are related to the ones in Ref.~\onlinecite{Abrikosov2002} by \(\Upsilon_{l,m}^{\pm}=V(\Omega)\,\mathcal{Y}_{lm}^{\,l\mp1/2}\), where \(V(\Omega)\) generates a local coordinate transformation such that \(V^\dagger(\Omega)\bigl(\hat{\mathbf{n}}(\Omega)\cdot\boldsymbol{\sigma}\bigr)V(\Omega)=\sigma_z\). 

In the remainder of the paper, we shall treat $j_{\rm max}$  as playing the role of finite system size. This is justified by the fact that the total number of single-particle states, $2(j_{\rm max}+1/2)(j_{\rm max}+3/2)$, is proportional to the number of spherical triangles of cutoff size $a$ on a sphere of unit radius.

We next introduce local interactions that preserve both the spatial \(\mathrm{SO(3)}\) symmetry and the discrete symmetries \(\mathcal{C}\), \(\mathcal{P}\) and \(\mathcal{T}\). Inspired by Eq.~(\ref{eq:Ham})~\footnote{Eq.~(\ref{eq:Ham}) is non-local, so we need to render it local and compatible with the spherical geometry.}, we write $H=H_0+H_{\rm int}$ with 
\beq \label{eq:Hint}
H_{\rm int}=- \int d\Omega_a d\Omega_b
U(\Omega_a-\Omega_b)
\psi_{\Omega_a}^\dagger
\sigma_{\Omega_a}
\psi_{\Omega_a}
\psi_{\Omega_b}^\dagger
\sigma_{\Omega_b}
\psi_{\Omega_b},\nn\\
\eeq
where \(\sigma_{\Omega} = \hat{\mathbf{n}}_{\Omega}\cdot\boldsymbol{\sigma}\) is the local spin operator aligned with the radial direction. Employing \(\sigma_{\Omega}\) ensures that the Hamiltonian commutes with the \(\mathrm{SO(3)}\) rotation, which also acts on the spin degrees of freedom via spin-orbit coupling in \(H_0\). Moreover, we avoid normal ordering so that charge-conjugation symmetry \(\mathcal{C}\) remains intact. Since \(U(\Omega_a-\Omega_b)\) must respect the same \(\mathrm{SO(3)}\) symmetry, it can be expanded in Haldane pseudopotentials~\cite{Zhu2023}:
\beq \label{eq:H}
U(\Omega_a-\Omega_b)
= {1\over 2}\sum_{l=0}^\infty
U_l\,
\nabla_\Omega^{(2l)}\,
\delta(\Omega_a-\Omega_b),
\eeq
whose matrix elements in the many-body basis involve Wigner \(3j\) symbols, as we detail in the Supplementary Material~\cite{supp}.

\textbf{Exact diagonalization.}
To perform exact diagonalization (ED), we construct the many-body Hilbert space from the single-particle orbitals \(\mathcal{Y}_{jm}^\ell\), retaining only those with \(j \le j_{\mathrm{max}}\). Because preserving the full \(\mathrm{SO(3)}\) symmetry requires complete angular momentum multiplets, the number of single-particle orbitals grows rapidly with \(j_{\mathrm{max}}\). For instance, \(j_{\mathrm{max}}=\tfrac72\) already yields 40 orbitals, leading to a many-body Hilbert space dimension that surpasses current ED capabilities. We therefore focus on \(j_{\mathrm{max}}=3/2\) and \(5/2\), giving two distinct system sizes for this study.

From mean-field theory, the transition between a time-reversal \(\mathcal{T}\)-symmetric (paramagnetic) phase and a \(\mathcal{T}\)-breaking (ferromagnetic) phase is primarily driven by the contact interaction \(U_0\). Fig.~\ref{fig2} (b) shows the order parameter \(\langle m^2\rangle\) versus \(U_0\), where
\beq
m=\int d\Omega\,\psi^\dagger(\Omega)\,\sigma_{\Omega}\,\psi(\Omega),
\eeq
is the mass operator on the sphere. We define $\tilde{m}=m/j_{\mathrm{max}}^{2-\Delta_m}$ and plot \(\langle\tilde{m}^2\rangle\) to compare among different system sizes. Limited by the available system size, we observe the \(\langle\tilde{m}^2\rangle\) curves intersect at \(U_0^c\sim 2\). 

\begin{figure*}[t]
  \centering
	\includegraphics[width=0.9\linewidth]{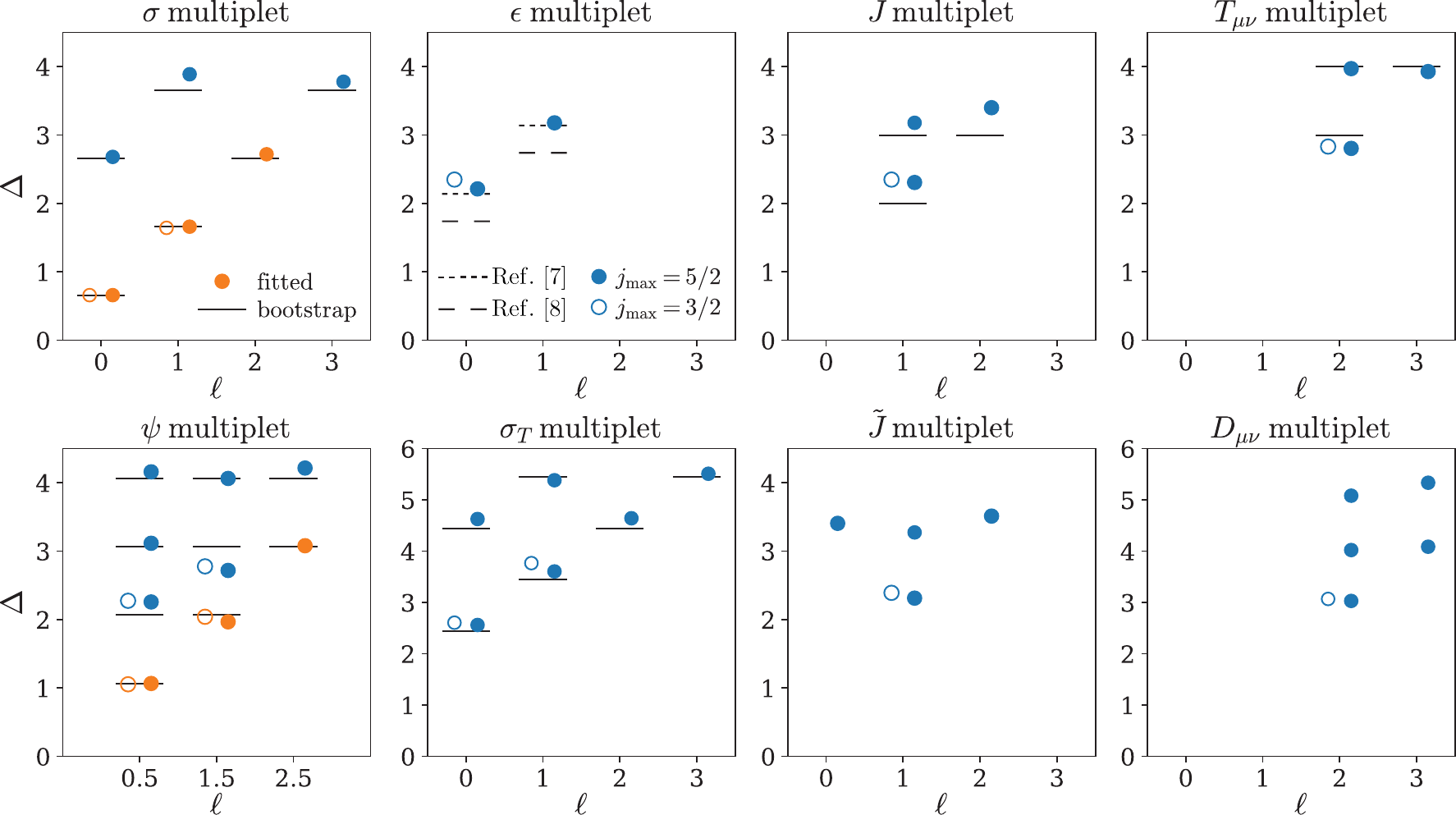}
	\caption{Comparison of conformal spectra from ED (dots) with conformal bootstrap predictions (lines). Each panel shows a distinct conformal multiplet. The horizontal axis denotes the total spin \(\ell\), and the vertical axis represents the scaling dimension \(\Delta\). ED points in orange are used to optimize the three pseudopotentials, whereas all blue points are predictions. Filled dots correspond to \(j_{\mathrm{max}}=5/2\), and open dots correspond to \(j_{\mathrm{max}}=3/2\). For \(\epsilon\sim\phi^2\), two sets of bootstrap predictions are shown: dashed lines include scalar correlators~\cite{Erramilli2023}, while dotted lines exclude them~\cite{Iliesiu2018}.}\label{fig:tower}
\end{figure*}

\textbf{Conformal spectrum.}
A central feature of conformal field theories (CFTs) is the \emph{state-operator correspondence}. 
We adopt radial quantization, which maps $\mathbb{R}^3$ to $\mathbb{S}^2 \times \mathbb{R}$, interpreting the radial coordinate as imaginary time and identifying the dilation operator with the Hamiltonian. 
At the conformal fixed point, each scaling operator corresponds to an energy eigenstate, with its scaling dimension determined by the properly normalized eigenenergy,
\begin{equation}
\delta E = E - E_0 = v\Delta,
\label{eq:Egap}
\end{equation}
where $E_0$ is the ground-state energy and \(v\) is a non-universal velocity. 
On $\mathbb{S}^2$ each operator is further labeled by its total spin \(\ell\) under \(\mathrm{SO(3)}\) (\(\ell \in \mathbb{Z}\) for bosons, \(\ell \in \mathbb{Z}+\tfrac12\) for fermions) and by a global \(U(1)\) charge \(q\). 
Although the parity symmetry $\mathcal{P}$ does not generally commute with \(\mathrm{SO(3)}\), singlet operators (\(\ell=0\)) retain a definite parity eigenvalue \(p_n\), and spin-$\tfrac12$ operators can split into \(m=\pm\tfrac12\) (here \(m\) is the eigenvalue of \(L_z\)) components of opposite parity. 
Therefore, each low-lying ED eigenstate \(|\mathcal{O}_n\rangle\) can be assigned \(\ell_n, q_n, \Delta_n\) and \(p_n\) whenever these quantum numbers are well defined. 
In the ED study we block-diagonalize the Hamiltonian in each sector of fixed \((q,m)\), and read off \(\ell\) from the familiar degeneracy pattern \(\{-\ell,-\ell+1,\dots,\ell\}\). 
Meanwhile, the eigenvalues \(E_n\) determine \(\Delta_n\) via Eq.~\eqref{eq:Egap}, up to the factor \(v\). 
Typically \(v\) is fixed by requiring the energy-momentum tensor to have its protected dimension \(\Delta_T=3\)~\cite{Zhu2023}; however, because high-scaling dimension operators are especially sensitive to finite-size and cutoff effects in our setup, we defer the calibration of \(v\) to the fitting procedure described below. 
Altogether, this procedure maps each eigenstate to an operator with quantum numbers \(\bigl(\ell_n,q_n,p_n\bigr)\) and approximate scaling dimension \(\Delta \approx \Delta_n\), enabling a systematic extraction of the conformal spectrum.

Although in the thermodynamic limit the multi-critical point can be reached by tuning a single coupling \(U_0\), the modest system sizes accessible here leave residual gaps that must be compensated by adjusting the next two pseudopotentials \(U_1\) and \(U_2\). At the same time, the availability of only two sizes precludes a reliable scaling-collapse analysis of the order parameter~\cite{Zhu2023}. The apparent crossing at \(U_0^c\) in Fig.~\ref{fig2} (b) should not be interpreted as a reliable estimate of the critical coupling. Rather, it only indicates that these limited system sizes capture the rapid onset of ferromagnetic order. Moreover, \(U_0^c\) follows the contact-only line $U_{1,2}=0$, whereas the spectral optimization uses $U_{1,2}$ as finite-cutoff counterterms, making the bare values of $U_0$ not directly comparable.
We therefore determine criticality spectroscopically: 
first, we fix \(v\) by matching the gap of the lowest scalar operator \(\sigma\) to its conformal-bootstrap value \(\Delta_\sigma=0.66\); 
with \(v\) in hand, we fine-tune \(\bigl(U_0,U_1,U_2\bigr)\) so that the scaling dimensions of the Dirac fermion \(\psi\), its lowest descendant $\partial_{\mu} \psi$, and the lowest descendant $\partial_{\mu} \sigma$ of \(\sigma\) agree with bootstrap results (orange dots in Fig.~\ref{fig:tower})~\footnote{To be precise, we minimize $\chi^2(U_0,U_1,U_2)=\sum_{\mathrm{Orange}~\mathcal{O}}|\Delta_\mathcal{O}(U_0,U_1,U_2)-\Delta_\mathcal{O}^\mathrm{Bootstrap}|^2$ for operators $\mathcal{O}$ marked by the orange dots.}. Here we use $\sigma$ instead of $T_{\mu\nu}$ to determine $v$, as $T_{\mu\nu}$ has a larger scaling dimension and is more sensitive to finite-size effect, especially given the modest system sizes. For \(j_{\mathrm{max}}=5/2\) this procedure yields \(\bigl(U_0,U_1,U_2\bigr) \approx (3.587, -0.068, -0.005)\); for \(j_{\mathrm{max}}=3/2\) we obtain \((4.854, -0.091, -0.027)\). We emphasize that all other operator scaling dimensions marked by blue dots in Fig.~\ref{fig:tower} are excluded from the optimization and therefore constitute genuine predictions, including the conformal towers of additional primaries $\tilde{J}_\mu$, $D_{\mu\nu}$.


Having pinpointed the critical point in parameter space, we find that the ED spectrum exhibits eight distinct \emph{conformal multiplets} spanning charge sectors \(q=0,1,2\), as shown in Fig.~\ref{fig:tower}. In the \(q=0\) sector, we identify the mass operator \(\sigma \sim \phi\), the scalar \(\epsilon \sim \phi^2\), the energy-momentum tensor \(T_{\mu\nu}\), and two currents: the conserved \(J_\mu \sim i\bar{\psi}\gamma_\mu\psi\) (satisfying \(\partial_\mu J_\mu=0\)) and a non-conserved current \(\tilde{J}_\mu \sim i\bar{\psi}\partial_\mu\psi - i\partial_\mu\bar{\psi}\psi\). For \(q=1\), the Dirac spinor \(\psi\) appears, while the \(q=2\) sector hosts two pairing operators: the singlet pairing  
\(\sigma_T \sim \psi^{\mathbf T}R\psi\)~\cite{Iliesiu2018} and a higher-spin \(\ell=2\) pairing \(D_{\mu\nu}\) (see Supplemental Material for details~\cite{supp}). Here \(R=i\sigma_y\) is the charge conjugation matrix.
Remarkably, the scaling dimension of \(D_{\mu\nu}\) lies very close to \(\Delta=3\), almost saturating the unitarity bound
\(\Delta=\ell+1\) for spin-\(\ell\) operators in \((2+1)\)-dimensional CFTs.  
This near-saturation, together with the absence of discernible descendants at \(\ell<2\) in our finite-size spectra, strongly suggests that \(D_{\mu\nu}\) acts as a \emph{nearly} conserved current.  
Such behavior accords with the weakly broken higher-spin symmetries expected in the GNY CFT, each accompanied by a nearly conserved tensor current~\cite{Giombi2017,Manashov2017,Zhou2023aa}.  
We conjecture that \(D_{\mu\nu}\) corresponds to an unconventional \(p\)-wave pairing operator  
\(
\psi^{\mathbf T}R(\gamma_{\mu}\partial_{\nu}+\gamma_{\nu}\partial_{\mu})\psi
\), 
which is the simplest conserved tensor beyond \(J_\mu\) and \(T_{\mu\nu}\) in the free fermion CFT, and lives in the correct charge and spin sector.  
Because previous studies focused almost exclusively on the \(q=0\) sector, this charged (\(q=2\)) current has until now remained undetected. The parity assignments for the \(\sigma\), \(\epsilon\), \(\sigma_T\), and \(\psi\) conformal multiplets all match theoretical expectations. Notably, \(\tilde{J}_\mu\) and \(D_{\mu\nu}\) were not covered by standard bootstrap or large-\(N\) expansions, rendering them new GNY data revealed by our finite-size ED approach.

Crucially, these multiplets provide direct evidence of emergent 3-dimensional conformal symmetry: each primary \(\mathcal{O}_{\mu_1\cdots\mu_\ell}\) generates the expected tower of descendants, verifying that the low-lying excitations indeed form representations of the special-conformal group. 
In particular, the two conserved operators \(J_\mu\) (\(\ell=1\)) and \(T_{\mu\nu}\) (\(\ell=2\)) exhibit \emph{restricted} descendant structures—because \(\partial_\mu J_\mu=0\) and \(\partial_\mu T_{\mu\nu}=0\) forbid lower-spin derivatives, their towers extend only to same or higher angular momentum. The nearly conserved operator \(D_{\mu\nu}\) shows an analogous pattern. By contrast, the non-conserved current $\tilde{J}_\mu$ has quantum number identical to that of $J_\mu$ but admits derivative descendants unavailable to $J_\mu$. The spinor multiplet displays its characteristic structure as well, where the Dirac contraction \(\slashed{\partial}\psi\equiv\gamma^\mu\partial_\mu\psi\) produces a same-spin descendant. Altogether, these observations confirm that our ED spectrum faithfully reproduces the conformal representation theory at the GNY critical point.


\setlength{\tabcolsep}{5.5pt}    
\renewcommand{\arraystretch}{1.4}  
\begin{table}[htbp]
\centering
\begin{tabular}{c |c| c |c| c| c| c| c}
\hline\hline
\(\mathcal{O}\)
 & \(\ell\)
 & \(q\)
 & \(\Delta_{3/2}\)
 & \(\Delta_{5/2}\)
 & \(\Delta_B\)
 & \(\text{Err}_{3/2}\)
 & \(\text{Err}_{5/2}\)
\\

\hline

\(\sigma\)
 & 0 & 0
 & 0.66 & 0.66
 & 0.66
 & -
 & -
\\

\(\partial_\mu\sigma\)
 & 1 & 0
 & 1.64 & 1.66
 & 1.66
 & 1.2\%
 & 0.0\%
\\

\(\partial^2\sigma\)
 & 0 & 0
 & 2.81 & 2.68
 & 2.66
 & 5.6\%
 & 0.8\%
\\

\(\partial_\mu\partial_\nu\sigma\)
 & 2 & 0
 & 2.75 & 2.71
 & 2.66
 & 3.4\%
 & 2.3\%
\\

\hline

\(\psi\)
 & 1/2 & 1
 & 1.05 & 1.07
 & 1.07
 & 1.9\%
 & 0.0\%
\\

\(\partial_\mu\psi\)
 & 3/2 & 1
 & 2.04 & 1.97
 & 2.07
 & 1.4\%
 & 4.8\%
\\

\(\partial_{\mu}\partial_{\nu}\psi\)
 & 5/2 & 1
 & / & 3.08
 & 3.07
 & /
 & 0.3\%
\\

\hline

\(\sigma_T\)
 & 0 & 2
 & 2.61 & 2.56
 & 2.45
 & 6.5\%
 & 4.5\%
\\

\(\partial_\mu\sigma_T\)
 & 1 & 2
 & 3.77 & 3.61
 & 3.45
 & 9.3\%
 & 4.6\%
\\

\hline

\multirow{2}{*}{\(\epsilon\)}
 & \multirow{2}{*}{0}
 & \multirow{2}{*}{0}
 & \multirow{2}{*}{2.35}
 & \multirow{2}{*}{2.21}
 & 1.74
 & 35\%
 & 27\%
\\
& & & & & 2.14
 & 9.8\%
 & 3.3\%
\\

\hline

\(J_\mu\)
 & 1 & 0
 & 2.35 & 2.31
 & 2.00
 & 17\%
 & 15\%
\\

\hline

\(T_{\mu\nu}\)
 & 2 & 0 
 & 2.83 & 2.81
 & 3.00
 & 5.7\%
 & 6.3\%
\\

\hline

\(\tilde{J}_\mu\)
 & 1 & 0
 & 2.39 & 2.31
 & /
 & /
 & /
\\

\hline

\(D_{\mu\nu}\)
 & 2 & 2
 & 3.07 & 3.03
 & /
 & /
 & /
\\

\hline\hline
\end{tabular}
\caption{%
Comparison of ED scaling dimensions (\(\Delta_{3/2}\), \(\Delta_{5/2}\)) at \(j_{\mathrm{max}}=3/2\) and \(5/2\) 
with conformal bootstrap values (\(\Delta_B\)). 
We show the total spin \(\ell\), charge \(q\), and scaling dimensions \(\Delta\) of each operator \(\mathcal{O}\). Two different bootstrap results for \(\epsilon\) are shown: \(1.74\) (includes scalar correlators \cite{Erramilli2023}) and \(2.14\) (excludes scalar correlators \cite{Iliesiu2018}). 
The ``Error'' columns give the percentage difference between \(\Delta_{\mathrm{ED}}\) and \(\Delta_B\).
}
\label{tab:Delta}
\end{table}

To quantify the accuracy of our operator dimensions, we compare the ED results for \(\sigma\), \(\psi\), and \(\sigma_T\), along with their lowest-lying descendants, to conformal bootstrap values in Table~\ref{tab:Delta}. Notably, we achieved this consistency by fitting only three parameters \(\bigl(U_0,U_1,U_2\bigr)\) to the leading scaling dimensions of \(\sigma\) and \(\psi\) multiplets; many subleading operators---such as \(\sigma_T\)---still come out within \(\sim3\%\) of the bootstrap predictions. We do observe larger discrepancies for the conserved operators \(J_\mu\) and \(T_{\mu\nu}\). The \(\sim 10\%\) deviations are likely attributable to finite size effect and the nonlocality introduced by the sharp angular-momentum cutoff: the scaling dimensions of $J_\mu$ and $T_{\mu\nu}$ are protected by local Ward identities, whereas the sharp angular-momentum cutoff makes the finite-size theory nonlocal. This protection can be restored only gradually with increasing cutoff. Without protection, there is no reason for $J_\mu$ and $T_{\mu\nu}$ being less sensitive to finite-size effects than generic operators. The effect is enhanced for $J_\mu$ due to the nearby nonconserved current $\tilde{J}_\mu$ that introduces stronger cutoff-induced mixing. This issue arises because the GNY theory contains both fermionic and bosonic critical degrees of freedom, which gives a denser low-energy spectrum and creates more opportunities for operators to mix at a fixed angular-momentum cutoff. We also note that \(\epsilon\sim\phi^2\) has two reported bootstrap values depending on whether scalar correlators are included~\cite{Iliesiu2018,Erramilli2023}; our result aligns with the version excluding them~\cite{Iliesiu2018}. Nevertheless, this agreement alone should not be interpreted as implying that the estimate of Ref.~\cite{Iliesiu2018} is intrinsically more accurate. Rather, our result provides an independent numerical benchmark for the discrepancy between the two bootstrap analyses, whose detailed resolution are left for future work. Despite these subtleties, the close match across most operators, especially $\sigma_T$ with a high scaling dimension, strongly validates our ED study and bolsters confidence in the two newly identified primaries, \(\tilde{J}_\mu\) and especially the nearly conserved pairing operator \(D_{\mu\nu}\), which were not captured by earlier theoretical analyses. Together, these findings underscore the utility of ED on spheres in uncovering the GNY CFT.

\textbf{Conclusions.} We have investigated the Gross-Neveu-Yukawa (GNY) universality class in \((2+1)\)-dimensional Dirac fermion systems by performing exact diagonalization (ED) on a sphere, where the curvature naturally circumvents the parity anomaly. By carefully tuning the contact interaction and additional Haldane pseudopotentials \(\bigl(U_0,U_1,U_2\bigr)\) with modest angular momentum cutoff $j_{\mathrm{max}}=3/2$ and $5/2$, we pinpoint an optimal parameter set that reproduces known operator scaling dimensions determined by the conformal bootstrap. 


Despite the finite-size limitations our results show that a robust GNY critical point can be realized in a modest Hilbert space. We identify eight distinct conformal multiplets whose tower structures, including the restricted descendant patterns of conserved and nearly conserved operators, provide direct evidence of emergent conformal symmetry. Furthermore, the low-lying operators such as the mass field \(\sigma\), the Dirac spinor \(\psi\), and the pairing field \(\sigma_T\) all match conformal bootstrap predictions to within a few percent, even at \(j_{\mathrm{max}}=3/2\). We further uncover two previously unreported primaries, $\tilde{J}_\mu$ and $D_{\mu\nu}$, thereby enriching the CFT data set for the GNY universality class. In particular, $D_{\mu\nu}$ represents, to our knowledge, the first \emph{numerical} evidence of weakly broken higher-spin symmetries, which had only been inferred from $\epsilon$- and large-$N$ expansions~\cite{Giombi2017,Manashov2017,Zhou2023aa}. These findings demonstrate that Dirac fermions on the sphere offer a powerful platform for investigating Dirac criticality and extracting detailed conformal data.

Looking ahead, it would be compelling to explore larger angular momentum cutoffs, additional fermion flavors, and to extract operator product expansion (OPE) coefficients to broaden the reach of spherical ED in studying Dirac and GNY criticalities. In particular, certain Dirac criticalities may exhibit supersymmetry—such as the \(N=1\) GNY CFT or the \(N=2\) Wess-Zumino model.

\textbf{Acknowledgments.} We thank Max Metlitski, Congjun Wu, Ning Su, Ruihua Fan, Yan-Qi Wang, and Hui Yang for helpful discussions. We thank Yin-Chen He especially for discussions on nearly conserved tensor currents and spinor descendants. ZQG and DHL were funded by the U.S. Department of Energy, Office of Science, Office of Basic Energy Sciences, Materials Sciences and Engineering Division under Contract No. DE-AC02-05-CH11231 (Theory of Materials program KC2301). This research uses the Lawrencium computational cluster provided by the Lawrence Berkeley National Laboratory (Supported by the U.S. Department of Energy, Office of Basic Energy Sciences under Contract No. DE-AC02-05-CH11231).

\bibliography{main}

\clearpage

\appendix
\section{End Matter}

\textbf{Mean-field analysis.} In the End Matter we supplement the examination of the Hamiltonian in Eq.~\eqref{eq:H} at the mean-field level, taking a contact interaction 
\(U(\Omega_a-\Omega_b)=\tfrac{u_0}{2}\,\delta(\Omega_a-\Omega_b)\) with \(u_0>0\). Defining the mass operator as  
\beq
m = \int d\Omega ~\psi^\dagger(\Omega)\,\sigma_z\,\psi(\Omega),
\eeq
we obtain the mean-field decomposition
\beq
H_{\mathrm{MF}} = H_0 - u_0\langle m\rangle\int d\Omega\,\psi^\dagger(\Omega)\sigma_\Omega\psi(\Omega).
\eeq
For \(u_0 < u_c\), $\langle m\rangle =0$ and the fermion remains gapless. Once \(u_0 > u_c\), a nonzero \(\langle m\rangle\) spontaneously breaks time-reversal symmetry, leading to a twofold degenerate ground state. An infinitesimal external field \(h\) in \(H_0\) then favors one of these two states, inducing a first-order transition between the topologically distinct phases at \(h<0\) or \(h>0\). At the mean-field level, this phase diagram is consistent with Fig.~\ref{fig1} in the main text.

To go beyond mean-field and capture the physics near the multicritical point, one can perform a Hubbard-Stratonovich transformation. The resulting Euclidean action is
\begin{equation}\label{eq:HS}
\begin{aligned}
S = \int d\tau\Bigl[
&\psi^\dagger(\Omega,\tau)\bigl(\partial_\tau + \phi(\Omega,\tau)\sigma_\Omega\bigr)\psi(\Omega,\tau)
\\
&+H_0\bigl[\psi(\Omega,\tau)\bigr]
+\frac{\phi(\Omega,\tau)^2}{2 u_0}
\Bigr],
\end{aligned}
\end{equation}
where the expectation value of the real scalar field \(\phi(\Omega,\tau)\) is the mean-field order parameter \(\langle \phi(\Omega,\tau)\rangle = m\). Fluctuations of \(\phi\) become increasingly important as the system approaches the multi-critical point. Identifying Eq.~\eqref{eq:HS} with the GNY action indicates that the multi-critical point of interacting Dirac fermions on the sphere in Eq.~\eqref{eq:H} belongs to the GNY universality class. 
Recent progress on the GNY CFT, via conformal bootstrap, \(\epsilon\)-expansions, large-\(N\) expansions, and QMC simulations, offers multiple avenues to benchmark the numerical results presented in the main text.

\clearpage

\onecolumngrid

\vspace{0.3cm}

\supplementarysection

\begin{center}
\Large{\bf Supplemental Material for ``Interacting Chern insulator transition on the sphere: revealing the Gross--Neveu--Yukawa criticality"}
\end{center}

\section{Wilson's regularization of a single-flavored massless Dirac fermion}

In this section we briefly review Wilson's approach to regularize a single-flavored massless Dirac fermion on square lattice. Consider a single-flavored massless Dirac fermion in $k$-space
\beq
H=\sum_{\mathbf{k}}\psi^\dagger_\mathbf{k}(-k_y\sigma_x+k_x\sigma_y)\psi_\mathbf{k}.
\eeq
This theory has a time reversal symmetry $T=i\sigma_y\mathcal{K}$ and a charge conjugation symmetry $C=\sigma_x\mathcal{K}$. To regularize it on a square lattice, periodicity in both $k_x$ and $k_y$ needs to be maintained, i.e. $k_x\sim k_x+2\pi$, $k_y\sim k_y+2\pi$. A naïve regularization reads
\beq
H_\mathrm{reg.}=\sum_{\mathbf{k}}\psi^\dagger_\mathbf{k}(-\sin{k_y}\sigma_x+\sin{k_x}\sigma_y)\psi_\mathbf{k},
\eeq
which remains gapless at $(k_x,k_y)=(0,0)$ and preserves both $T$ and $C$ symmetries. However, this regularization introduces spurious Dirac nodes at $(k_x,k_y)=(0,\pi)$, $(\pi,0)$, and $(\pi,\pi)$. To gap out these extra Dirac nodes, Wilson proposed a $k$-dependent mass term, and the regularized Hamiltonian becomes
\beq
H_\mathrm{Wilson}=\sum_{\mathbf{k}}\psi^\dagger_\mathbf{k}[-\sin{k_y}\sigma_x+\sin{k_x}\sigma_y+(2-\cos{k_x}-\cos{k_y})m\sigma_z]\psi_\mathbf{k}.\label{eq:HWil}
\eeq
Though \Eq{eq:HWil} is gapless only at $(k_x,k_y)=(0,0)$, it explicitly breaks the time reversal symmetry $T$. Thus, to simulate a single-flavored massless Dirac fermion via Wilson's regularization, one inevitable needs rather large system size to have the time reversal symmetry emerge in the low energy, which could be painful in numerics.

\section{Dirac fermion on the sphere}

\subsection{Coordinate dependent approach}

In this section we use $X$, $Y$, $Z$, and $I$ as shorthands for Pauli Matrices $\sigma_x$, $\sigma_y$, $\sigma_z$, and $\sigma_0$, respectively. For an $N$-flavor Dirac fermion $\psi(\Omega)$ on a sphere, the free part of the Hamiltonian reads
\bea
H_0=\int\mathrm{d}\Omega~\psi^\dagger(\Omega)(-i\nabla_\Omega\otimes I_N)\psi(\Omega)+h\psi^\dagger(\Omega)(Z\otimes M)\psi(\Omega),\label{eq:D}
\eea
where $I_N$ is an $N$ by $N$ identity matrix, and $M$ is the $N$ by $N$ real Hermitian mass matrix with $M^2=I_N$. The Dirac fermion 
$\psi(\Omega)=(\psi_+(\Omega), \psi_-(\Omega))^\mathbf{T}$ is defined in spherical coordinate $\Omega=(\theta,\varphi)$, and
\beq
-i\nabla_\Omega=-\frac{iX}{R}\left(\partial_\theta+\frac{\cot{\theta}}{2}\right)-\frac{iY}{R\sin{\theta}}\partial_\varphi\label{eq:kernel}
\eeq
is the Dirac operator on the sphere with radius $R$, which has a time reversal symmetry $T=iY\mathcal{K}$ with $T^2=-I$ protecting the gaplessness of its spectrum and a charge conjugation symmetry $C=X\mathcal{K}$, such that $T(-i\nabla_\Omega)T^{-1}=-C(-i\nabla_\Omega)C^{-1}=(-i\nabla_\Omega)$. It also possesses a parity symmetry $P=Y$ with $\theta\mapsto\pi-\theta$, $\varphi\mapsto\varphi$. The mass term, with coupling constant $h$, explicitly breaks the time-reversal symmetry $T$ while preserving charge conjugation symmetry $C$. Hereafter, we set $h=0$ by default to ensure that time-reversal symmetry remains unbroken.

The eigenfunctions of $-i\nabla_\Omega$ are the two-component spinor spherical functions $\Upsilon_{s,l,m}(\Omega)$ satisfying
\beq
-i\nabla_\Omega\Upsilon_{s,l,m}(\Omega)=s E_{l,m}\Upsilon_{s,l,m}(\Omega),
\eeq
where $s =\pm 1$, $l$ is the angular momentum, and $m=-l,(-l+1),...,-1/2,1/2,...,(l-1),l$. Here both $l$ and $m$ are half-odd numbers to satisfy the anti-periodic boundary condition of the fermion.
The eigen kinetic energy is given by
$E_{l,m}=(l+1/2)/R$. It is linear in $l$ and become gapless as $R\rightarrow\infty$.  Setting the ultraviolet cutoff at $l\le l_\mathrm{Max}$ allows the fermion operator to be mode-expanded as
\beq
\psi(\Omega)=\sum_{l=1/2}^{l_\mathrm{Max}}\sum_{m=-l}^l\left(\Upsilon_{+,l,m}(\Omega),\Upsilon_{-,l,m}(\Omega)\right) F_{l,m},\label{eq:mode}
\eeq
with 
\beq
F_{l,m}^\dagger =(f^\dagger_{+,l,m},f^\dagger_{-,l,m})^\mathbf{T}=\int\mathrm{d}\Omega~\psi(\Omega)^\dagger\left(\Upsilon_{+,l,m}(\Omega),\Upsilon_{-,l,m}(\Omega)\right).
\eeq
As usual, the orthogonality of $\Upsilon_{s,l,m}(\Omega)$
ensures the anti-commutation relation of $F_{l,m}$. Thus $H_0$ is diagonalized as:
\beq
H_0=\sum_{l,m}\frac{2l+1}{2R}\big(f_{+,l,m}^\dagger f_{+,l,m}-f_{-,l,m}^\dagger f_{-,l,m}\big)=\sum_{l,m}\frac{2l+1}{2R}F^\dagger_{l,m}[Z\otimes I_N] F^{l,m}.\label{eq:H_0}
\eeq 

The discrete symmetries act on the Dirac fermion as
\beq
&&\mathcal{C}\psi^\dagger(\Omega)\mathcal{C}^{-1}=\psi^\mathbf{T}(\Omega)(X\otimes I_N),\\
&&\mathcal{T}\psi^\dagger(\Omega)\mathcal{T}^{-1}=\psi^\dagger(\Omega)(-iY\otimes I_N),~~\mathcal{T}i\mathcal{T}^{-1}=-i,\\
&&\mathcal{P}\psi^\dagger(\theta,\varphi)\mathcal{P}^{-1}=\psi^\dagger(\pi-\theta,\varphi)(Y\otimes I_N),
\eeq
where $\mathcal{C}$, $\mathcal{T}$, and $\mathcal{P}$ stands for charge conjugation, time reversal, and parity transformations. Their actions on $F_{l,m}$ basis is derived below in details.
\beq
\mathcal{C}F^\dagger_{l,m}\mathcal{C}^{-1}&=&\int\mathrm{d}\Omega~\psi^\mathbf{T}(\Omega)\left(X\Upsilon_{+,l,m}(\Omega),X\Upsilon_{-,l,m}(\Omega)\right)\otimes I_N,\nn\\
&=&\int\mathrm{d}\Omega~\psi^\mathbf{T}(\Omega)\left(Y\Upsilon_{-,l,m}(\Omega),-Y\Upsilon_{+,l,m}(\Omega)\right)\otimes I_N\nn\\
&=&(-1)^{l-m}\int\mathrm{d}\Omega~\psi^\mathbf{T}(\Omega)\left(i\Upsilon^*_{-,l,-m}(\Omega),-i\Upsilon^*_{+,l,-m}(\Omega)\right)\otimes I_N\nn\\
&=&(-1)^{l-m}\int\mathrm{d}\Omega~\left[\begin{pmatrix}
    i\Upsilon^\dagger_{-,l,-m}(\Omega) \\ -i\Upsilon^\dagger_{+,l,-m}(\Omega)
\end{pmatrix}\otimes I_N \psi(\Omega)\right]^\mathbf{T}\nn\\
&=&(-1)^{l-m}\left[-Y\otimes I_N\int\mathrm{d}\Omega~\begin{pmatrix}
    \Upsilon^\dagger_{+,l,-m}(\Omega) \\ \Upsilon^\dagger_{-,l,-m}(\Omega)
\end{pmatrix}\otimes I_N \psi(\Omega)\right]^\mathbf{T}\nn\\
&=&(-1)^{l-m} F_{l,-m}^\mathbf{T}(Y\otimes I_N),\\
\mathcal{T}F^\dagger_{l,m}\mathcal{T}^{-1}&=&\int\mathrm{d}\Omega~\psi^\dagger(\Omega)\left(-iY\Upsilon^*_{+,l,m}(\Omega),-iY\Upsilon^*_{-,l,m}(\Omega)\right)\otimes I_N\nn\\
&=&(-1)^{l-m} \int\mathrm{d}\Omega~\psi^\dagger(\Omega)\left(\Upsilon_{+,l,-m}(\Omega),\Upsilon_{-,l,-m}(\Omega)\right)\otimes I_N\nn\\
&=&(-1)^{l-m}F^\dagger_{l,-m},\\
\mathcal{P}F^\dagger_{l,m}\mathcal{P}^{-1}&=&\int\mathrm{d}\Omega~\psi^\dagger(\pi-\theta,\varphi)\left(Y\Upsilon_{+,l,m}(\Omega),Y\Upsilon_{-,l,m}(\Omega)\right)\otimes I_N\nn\\
&=&-(-1)^{l-m}\int\mathrm{d}\Omega~\psi^\dagger(\Omega)\left(\Upsilon_{+,l,-m}(\Omega),-\Upsilon_{-,l,m}(\Omega)\right)\otimes I_N\nn\\
&=&-(-1)^{l-m}F^\dagger_{l,m}(Z\otimes I_N),
\eeq
where $-iY\Upsilon_{s,l,m}=(-1)^{l-m}\Upsilon^*_{s,l,-m}$, $Z\Upsilon_{s,l,m}=is\Upsilon_{-s,l,m}$, and $\Upsilon_{s,l,m}(\pi-\theta,\varphi)=-s(-1)^{l-m}Y\Upsilon_{s,l,m}(\theta,\varphi)$ are used. Similarly,
\beq
\mathcal{T}F_{l,m}\mathcal{T}^{-1}=(-1)^{l-m}F_{l,-m}, ~~\mathcal{C}F_{l,m}\mathcal{C}^{-1}=(-1)^{l-m}(Y\otimes I_N)(F^\dagger_{l,-m})^\mathbf{T},~~\mathcal{P}F_{l,m}\mathcal{P}^{-1}=-(-1)^{l-m}(Z\otimes I_N)F_{l,m}.\nn
\eeq

A generalized interaction Hamiltonian with both $ZZ$- and $II$-type interactions read
\beq
H_\mathrm{int}=
&-&\int\mathrm{d}\Omega_a\mathrm{d}\Omega_b~U(\Omega_a-\Omega_b)\psi^\dagger(\Omega_a)(Z\otimes M)\psi(\Omega_a)\psi^\dagger(\Omega_b)(Z\otimes M)\psi(\Omega_b)\nn\\
&-&\int\mathrm{d}\Omega_a\mathrm{d}\Omega_b~U^\prime(\Omega_a-\Omega_b)\psi^\dagger(\Omega_a)(I\otimes Q)\psi(\Omega_a)\psi^\dagger(\Omega_b)(I\otimes Q)\psi(\Omega_b),
\eeq
where $Q$ is an $N$ by $N$ Hermitian matrix with $Q^2=I_N$.
Let us first look at the term $\psi^\dagger(\Omega_a)(Z\otimes M)\psi(\Omega_a)$ and $\psi^\dagger(\Omega_a)(I\otimes Q)\psi(\Omega_a)$. It can be mode expanded in terms of $f_{s,l,m}$ as
\beq
&&\psi^\dagger(\Omega_a)(Z\otimes M)\psi(\Omega_a)\nn\\
&=&\sum_{l_1,l_2,m_1,m_2}\left(f^\dagger_{+,l_1,m_1},f^\dagger_{-,l_1,m_1}\right)
\left[\begin{pmatrix}
    \Upsilon^\dagger_{+,l_1,m_1} \\ \Upsilon^\dagger_{-,l_1,m_1}
\end{pmatrix}\otimes I_N \right](Z\otimes M)\left[\left(\Upsilon_{+,l_2,m_2},\Upsilon_{-,l_2,m_2}\right)\otimes I_N\right]
\begin{pmatrix}
    f_{+,l_2,m_2} \\ f_{-,l_2,m_2}
\end{pmatrix}\nn\\
&=&\sum_{l_1,l_2,m_1,m_2}\left(f^\dagger_{+,l_1,m_1},f^\dagger_{-,l_1,m_1}\right)
\left[\begin{pmatrix}
    \Upsilon^\dagger_{+,l_1,m_1} \\ \Upsilon^\dagger_{-,l_1,m_1}
\end{pmatrix}Z\left(\Upsilon_{+,l_2,m_2},\Upsilon_{-,l_2,m_2}\right)\otimes M\right]
\begin{pmatrix}
    f_{+,l_2,m_2} \\ f_{-,l_2,m_2}
\end{pmatrix}\nn\\
&=&
i\sum_{l_1,l_2,m_1,m_2}\left(f^\dagger_{+,l_1,m_1},f^\dagger_{-,l_1,m_1}\right)
\left[\begin{pmatrix}
    \Upsilon^\dagger_{+,l_1,m_1}\Upsilon_{-,l_2,m_2} & -\Upsilon^\dagger_{+,l_1,m_1}\Upsilon_{+,l_2,m_2} \\
    \Upsilon^\dagger_{-,l_1,m_1}\Upsilon_{-,l_2,m_2} & -\Upsilon^\dagger_{-,l_1,m_1}\Upsilon_{+,l_2,m_2}
\end{pmatrix}\otimes M\right]
\begin{pmatrix}
    f_{+,l_2,m_2} \\ f_{-,l_2,m_2}
\end{pmatrix}\nn\\
&=&i\sum_{s_1,s_2,l_1,l_2,m_1,m_2}f^\dagger_{s_1,l_1,m_1}\left(s_2\Upsilon^\dagger_{s_1,l_1,m_1}\Upsilon_{-s_2,l_2,m_2}\right)M f^\dagger_{s_2,l_2,m_2},\label{eq:Zright}\\
&&\psi^\dagger(\Omega_a)(I\otimes Q)\psi(\Omega_a)\nn\\
&=&\sum_{l_1,l_2,m_1,m_2}\left(f^\dagger_{+,l_1,m_1},f^\dagger_{-,l_1,m_1}\right)
\left[\begin{pmatrix}
    \Upsilon^\dagger_{+,l_1,m_1}\Upsilon_{+,l_2,m_2} & \Upsilon^\dagger_{+,l_1,m_1}\Upsilon_{-,l_2,m_2} \\
    \Upsilon^\dagger_{-,l_1,m_1}\Upsilon_{+,l_2,m_2} & \Upsilon^\dagger_{-,l_1,m_1}\Upsilon_{-,l_2,m_2}
\end{pmatrix}\otimes Q\right]
\begin{pmatrix}
    f_{+,l_2,m_2} \\ f_{-,l_2,m_2}
\end{pmatrix}\nn\\
&=&\sum_{s_1,s_2,l_1,l_2,m_1,m_2}f^\dagger_{s_1,l_1,m_1}\left(\Upsilon^\dagger_{s_1,l_1,m_1}\Upsilon_{s_2,l_2,m_2}\right)M f^\dagger_{s_2,l_2,m_2}
\eeq
where in passing to the last line of each equality we use $Z\Upsilon_{s,l,m}=is\Upsilon_{-s,l,m}$. For clarity we omit the $\Omega_a$ dependence from the second line. Note that the Pauli matrix $Z$ can also act leftward to $(\Upsilon^\dagger,\Upsilon^\dagger)^\mathbf{T}$ as $\Upsilon^\dagger_{s,l,m}Z=-is\Upsilon^\dagger_{-s,l,m}$. This yields
\beq
&&\psi^\dagger(\Omega_a)(Z\otimes M)\psi(\Omega_a)\nn\\
&=&
i\sum_{l_1,l_2,m_1,m_2}\left(f^\dagger_{+,l_1,m_1},f^\dagger_{-,l_1,m_1}\right)
\left[\begin{pmatrix}
    -\Upsilon^\dagger_{-,l_1,m_1}\Upsilon_{+,l_2,m_2} & -\Upsilon^\dagger_{-,l_1,m_1}\Upsilon_{-,l_2,m_2} \\
    \Upsilon^\dagger_{+,l_1,m_1}\Upsilon_{+,l_2,m_2} & \Upsilon^\dagger_{+,l_1,m_1}\Upsilon_{-,l_2,m_2}
\end{pmatrix}\otimes M\right]
\begin{pmatrix}
    f_{+,l_2,m_2} \\ f_{-,l_2,m_2}
\end{pmatrix}\nn\\
&=&i\sum_{s_1,s_2,l_1,l_2,m_1,m_2}f^\dagger_{s_1,l_1,m_1}\left(-s_1\Upsilon^\dagger_{-s_1,l_1,m_1}\Upsilon_{s_2,l_2,m_2}\right)M f^\dagger_{s_2,l_2,m_2}.\label{eq:Zleft}
\eeq
Thus equality
\beq
s_2\Upsilon^\dagger_{s_1,l_1,m_1}\Upsilon_{-s_2,l_2,m_2}+s_1\Upsilon^\dagger_{-s_1,l_1,m_1}\Upsilon_{s_2,l_2,m_2}=0.\label{eq:check}
\eeq
should always hold to ensure \Eq{eq:Zleft} equal to \Eq{eq:Zright}. In fact this is also the Hermitian condition of $\psi^\dagger(\Omega_a)(Z\otimes M)\psi(\Omega_a)$. According to the orthogonality relation
\beq
\int\mathrm{d}\Omega~\Upsilon^\dagger_{s^\prime,l^\prime,m^\prime}(\Omega)\Upsilon_{s,l,m}(\Omega)=\delta_{ss^\prime}\delta_{ll^\prime}\delta_{mm^\prime},\label{eq:ortho}
\eeq 
the mass term and the chemical potential term read
\beq
&&\int\mathrm{d}\Omega~\psi^\dagger(\Omega)(Z\otimes M)\psi(\Omega)=\sum_{l,m}F^\dagger_{l,m}(Y\otimes M) F_{l,m},\\
&&\int\mathrm{d}\Omega~\psi^\dagger(\Omega)(I\otimes Q)\psi(\Omega)=\sum_{l,m}F^\dagger_{l,m}(I\otimes Q) F_{l,m},
\eeq
where the mass term $Y\otimes M$ indeed opens a gap.

The spinor spherical function is related to spherical harmonics via
\beq
\Upsilon_{s_1,l_1,m_1}(\Omega)=W(\Omega)
\begin{pmatrix}
    i^{l_1-\frac{1}{2}}\sqrt{\frac{l_1+m_1}{2l_1}}Y_{l_1-\frac{1}{2},m_1-\frac{1}{2}}(\Omega)+i^{l_1+\frac{1}{2}}s_1\sqrt{\frac{l_1-m_1+1}{2l_1+2}}Y_{l_1+\frac{1}{2},m_1-\frac{1}{2}}(\Omega)\\
    i^{l_1-\frac{1}{2}}\sqrt{\frac{l_1-m_1}{2l_1}}Y_{l_1-\frac{1}{2},m_1+\frac{1}{2}}(\Omega)-i^{l_1+\frac{1}{2}}s_1\sqrt{\frac{l_1+m_1+1}{2l_1+2}}Y_{l_1+\frac{1}{2},m_1+\frac{1}{2}}(\Omega)
\end{pmatrix},
\eeq
where $W(\Omega)=\exp{(iY\theta/2)}\exp{(iZ\varphi/2)}$. The inner product between two spinor spherical functions can be explicitly computed as
\beq
&&\Upsilon^\dagger_{s_1,l_1,m_1}\Upsilon_{s_2,l_2,m_2}\nn\\
&=&
\begin{pmatrix}
    (-i)^{l_1-\frac{1}{2}}\sqrt{\frac{l_1+m_1}{2l_1}}Y^*_{l_1-\frac{1}{2},m_1-\frac{1}{2}}+(-i)^{l_1+\frac{1}{2}}s_1\sqrt{\frac{l_1-m_1+1}{2l_1+2}}Y^*_{l_1+\frac{1}{2},m_1-\frac{1}{2}}\\
    (-i)^{l_1-\frac{1}{2}}\sqrt{\frac{l_1-m_1}{2l_1}}Y^*_{l_1-\frac{1}{2},m_1+\frac{1}{2}}-(-i)^{l_1+\frac{1}{2}}s_1\sqrt{\frac{l_1+m_1+1}{2l_1+2}}Y^*_{l_1+\frac{1}{2},m_1+\frac{1}{2}}
\end{pmatrix}
W^\dagger\cdot\nn\\
&& W
\begin{pmatrix}
    i^{l_2-\frac{1}{2}}\sqrt{\frac{l_2+m_2}{2l_2}}Y_{l_2-\frac{1}{2},m_2-\frac{1}{2}}+i^{l_2+\frac{1}{2}}s_2 \sqrt{\frac{l_2-m_2+1}{2l_2+2}}Y_{l_2+\frac{1}{2},m_2-\frac{1}{2}}\\
    i^{l_2-\frac{1}{2}}\sqrt{\frac{l_2-m_2}{2l_2}}Y_{l_2-\frac{1}{2},m_2+\frac{1}{2}}-i^{l_2+\frac{1}{2}}s_2 \sqrt{\frac{l_2+m_2+1}{2l_2+2}}Y_{l_2+\frac{1}{2},m_2+\frac{1}{2}}
\end{pmatrix}^\mathbf{T}\nn\\
&=&\Bigg[\sqrt{\frac{l_1+m_1}{2l_1}\frac{l_2+m_2}{2l_2}}Y^*_{l_1-\frac{1}{2},m_1-\frac{1}{2}}Y_{l_2-\frac{1}{2},m_2-\frac{1}{2}}+\sqrt{\frac{l_1-m_1}{2l_1}\frac{l_2-m_2}{2l_2}}Y^*_{l_1-\frac{1}{2},m_1+\frac{1}{2}}Y_{l_2-\frac{1}{2},m_2+\frac{1}{2}}\nn\\
& &\quad -is_1\Big(\sqrt{\frac{l_1-m_1+1}{2l_1+2}\frac{l_2+m_2}{2l_2}}Y^*_{l_1+\frac{1}{2},m_1-\frac{1}{2}}Y_{l_2-\frac{1}{2},m_2-\frac{1}{2}}-\sqrt{\frac{l_1+m_1+1}{2l_1+2}\frac{l_2-m_2}{2l_2}}Y^*_{l_1+\frac{1}{2},m_1+\frac{1}{2}}Y_{l_2-\frac{1}{2},m_2+\frac{1}{2}}\Big)\nn\\
& &\quad +is_2\Big(\sqrt{\frac{l_1+m_1}{2l_1}\frac{l_2-m_2+1}{2l_2+2}}Y^*_{l_1-\frac{1}{2},m_1-\frac{1}{2}}Y_{l_2+\frac{1}{2},m_2-\frac{1}{2}}-\sqrt{\frac{l_1-m_1}{2l_1}\frac{l_2+m_2+1}{2l_2+2}}Y^*_{l_1-\frac{1}{2},m_1+\frac{1}{2}}Y_{l_2+\frac{1}{2},m_2+\frac{1}{2}}\Big)\nn\\
& &\quad +s_1s_2\Big(\sqrt{\frac{l_1-m_1+1}{2l_1+2}\frac{l_2-m_2+1}{2l_2+2}}Y^*_{l_1+\frac{1}{2},m_1-\frac{1}{2}}Y_{l_2+\frac{1}{2},m_2-\frac{1}{2}}+\sqrt{\frac{l_1+m_1+1}{2l_1+2}\frac{l_2+m_2+1}{2l_2+2}}Y^*_{l_1+\frac{1}{2},m_1+\frac{1}{2}}Y_{l_2+\frac{1}{2},m_2+\frac{1}{2}}\Big)\Bigg]\nn\\
& &\times i^{l_2-l_1}.\label{eq:YY}
\eeq
The Hermitian check \Eq{eq:check} is automatically satisfied by \Eq{eq:YY}. By means of
\beq
\int\mathrm{d}\Omega ~Y_{j_1,\nu_1}(\Omega)Y_{j_2,\nu_2}(\Omega)Y_{j_3,\nu_3}(\Omega)=\sqrt{\frac{(2j_1+1)(2j_2+1)(2j_3+1)}{4\pi}}
\begin{pmatrix}
    j_1 & j_2 & j_3 \\ 0 & 0 & 0
\end{pmatrix}
\begin{pmatrix}
    j_1 & j_2 & j_3 \\ \nu_1 & \nu_2 & \nu_3
\end{pmatrix},
\eeq
and $Y_{j,\nu}^*(\Omega)=(-1)^\nu Y_{j,-\nu}(\Omega)$, the integration over solid angle $\Omega_a$ ($\Omega_b$) can be exactly performed as
\beq
&&\sqrt{\frac{4\pi}{2k+1}}\int\mathrm{d}\Omega_a ~Y^*_{k,n}(\Omega_a)\Upsilon^\dagger_{s_1,l_1,m_1}(\Omega_a)\Upsilon_{s_2,l_2,m_2}(\Omega_a)\nn\\
&=&i^{l_2-l_1}(-1)^{n+m_1-\frac{1}{2}}\sqrt{(l_1+m_1)(l_2+m_2)}
\begin{pmatrix}
    k & l_1-\frac{1}{2} & l_2-\frac{1}{2} \\ 0 & 0 & 0
\end{pmatrix}
\begin{pmatrix}
    k & l_1-\frac{1}{2} & l_2-\frac{1}{2} \\ -n & -m_1+\frac{1}{2} & m_2-\frac{1}{2}
\end{pmatrix}\nn\\
&\quad &+i^{l_2-l_1}(-1)^{n+m_1+\frac{1}{2}}\sqrt{(l_1-m_1)(l_2-m_2)}
\begin{pmatrix}
    k & l_1-\frac{1}{2} & l_2-\frac{1}{2} \\ 0 & 0 & 0
\end{pmatrix}
\begin{pmatrix}
    k & l_1-\frac{1}{2} & l_2-\frac{1}{2} \\ -n & -m_1-\frac{1}{2} & m_2+\frac{1}{2}
\end{pmatrix}\nn\\
&\quad &+i^{l_2-l_1-1}(-1)^{n+m_1-\frac{1}{2}}s_1\sqrt{(l_1-m_1+1)(l_2+m_2)}
\begin{pmatrix}
    k & l_1+\frac{1}{2} & l_2-\frac{1}{2} \\ 0 & 0 & 0
\end{pmatrix}
\begin{pmatrix}
    k & l_1+\frac{1}{2} & l_2-\frac{1}{2} \\ -n & -m_1+\frac{1}{2} & m_2-\frac{1}{2}
\end{pmatrix}\nn\\
&\quad &-i^{l_2-l_1-1}(-1)^{n+m_1+\frac{1}{2}}s_1\sqrt{(l_1+m_1+1)(l_2-m_2)}
\begin{pmatrix}
    k & l_1+\frac{1}{2} & l_2-\frac{1}{2} \\ 0 & 0 & 0
\end{pmatrix}
\begin{pmatrix}
    k & l_1+\frac{1}{2} & l_2-\frac{1}{2} \\ -n & -m_1-\frac{1}{2} & m_2+\frac{1}{2}
\end{pmatrix}\nn\\
&\quad &+i^{l_2-l_1+1}(-1)^{n+m_1-\frac{1}{2}}s_2\sqrt{(l_1+m_1)(l_2-m_2+1)}
\begin{pmatrix}
    k & l_1-\frac{1}{2} & l_2+\frac{1}{2} \\ 0 & 0 & 0
\end{pmatrix}
\begin{pmatrix}
    k & l_1-\frac{1}{2} & l_2+\frac{1}{2} \\ -n & -m_1+\frac{1}{2} & m_2-\frac{1}{2}
\end{pmatrix}\nn\\
&\quad &-i^{l_2-l_1+1}(-1)^{n+m_1+\frac{1}{2}}s_2\sqrt{(l_1-m_1)(l_2+m_2+1)}
\begin{pmatrix}
    k & l_1-\frac{1}{2} & l_2+\frac{1}{2} \\ 0 & 0 & 0
\end{pmatrix}
\begin{pmatrix}
    k & l_1-\frac{1}{2} & l_2+\frac{1}{2} \\ -n & -m_1-\frac{1}{2} & m_2+\frac{1}{2}
\end{pmatrix}\nn\\
&\quad &+i^{l_2-l_1}(-1)^{n+m_1-\frac{1}{2}}s_1s_2\sqrt{(l_1-m_1+1)(l_2-m_2+1)}
\begin{pmatrix}
    k & l_1+\frac{1}{2} & l_2+\frac{1}{2} \\ 0 & 0 & 0
\end{pmatrix}
\begin{pmatrix}
    k & l_1+\frac{1}{2} & l_2+\frac{1}{2} \\ -n & -m_1+\frac{1}{2} & m_2-\frac{1}{2}
\end{pmatrix}\nn\\
&\quad &+i^{l_2-l_1}(-1)^{n+m_1+\frac{1}{2}}s_1s_2\sqrt{(l_1+m_1+1)(l_2+m_2+1)}
\begin{pmatrix}
    k & l_1+\frac{1}{2} & l_2+\frac{1}{2} \\ 0 & 0 & 0
\end{pmatrix}
\begin{pmatrix}
    k & l_1+\frac{1}{2} & l_2+\frac{1}{2} \\ -n & -m_1-\frac{1}{2} & m_2+\frac{1}{2}
\end{pmatrix},
\eeq
and
\beq
&&\sqrt{\frac{4\pi}{2k+1}}\int\mathrm{d}\Omega_b ~Y_{k,n}(\Omega_b)\Upsilon^\dagger_{s_3,l_3,m_3}(\Omega_b)\Upsilon_{s_4,l_4,m_4}(\Omega_b)\nn\\
&=&i^{l_4-l_3}(-1)^{m_3-\frac{1}{2}}\sqrt{(l_3+m_3)(l_4+m_4)}
\begin{pmatrix}
    k & l_3-\frac{1}{2} & l_4-\frac{1}{2} \\ 0 & 0 & 0
\end{pmatrix}
\begin{pmatrix}
    k & l_3-\frac{1}{2} & l_4-\frac{1}{2} \\ n & -m_3+\frac{1}{2} & m_4-\frac{1}{2}
\end{pmatrix}\nn\\
&\quad &+i^{l_4-l_3}(-1)^{m_3+\frac{1}{2}}\sqrt{(l_3-m_3)(l_4-m_4)}
\begin{pmatrix}
    k & l_3-\frac{1}{2} & l_4-\frac{1}{2} \\ 0 & 0 & 0
\end{pmatrix}
\begin{pmatrix}
    k & l_3-\frac{1}{2} & l_4-\frac{1}{2} \\ n & -m_3-\frac{1}{2} & m_4+\frac{1}{2}
\end{pmatrix}\nn\\
&\quad &+i^{l_4-l_3-1}(-1)^{m_3-\frac{1}{2}}s_3\sqrt{(l_3-m_3+1)(l_4+m_4)}
\begin{pmatrix}
    k & l_3+\frac{1}{2} & l_4-\frac{1}{2} \\ 0 & 0 & 0
\end{pmatrix}
\begin{pmatrix}
    k & l_3+\frac{1}{2} & l_4-\frac{1}{2} \\ n & -m_3+\frac{1}{2} & m_4-\frac{1}{2}
\end{pmatrix}\nn\\
&\quad &-i^{l_4-l_3-1}(-1)^{m_3+\frac{1}{2}}s_3\sqrt{(l_3+m_3+1)(l_4-m_4)}
\begin{pmatrix}
    k & l_3+\frac{1}{2} & l_4-\frac{1}{2} \\ 0 & 0 & 0
\end{pmatrix}
\begin{pmatrix}
    k & l_3+\frac{1}{2} & l_4-\frac{1}{2} \\ n & -m_3-\frac{1}{2} & m_4+\frac{1}{2}
\end{pmatrix}\nn\\
&\quad &+i^{l_4-l_3+1}(-1)^{m_3-\frac{1}{2}}s_4\sqrt{(l_3+m_3)(l_4-m_4+1)}
\begin{pmatrix}
    k & l_3-\frac{1}{2} & l_4+\frac{1}{2} \\ 0 & 0 & 0
\end{pmatrix}
\begin{pmatrix}
    k & l_3-\frac{1}{2} & l_4+\frac{1}{2} \\ n & -m_3+\frac{1}{2} & m_4-\frac{1}{2}
\end{pmatrix}\nn\\
&\quad &-i^{l_4-l_3+1}(-1)^{m_3+\frac{1}{2}}s_4\sqrt{(l_3-m_3)(l_4+m_4+1)}
\begin{pmatrix}
    k & l_3-\frac{1}{2} & l_4+\frac{1}{2} \\ 0 & 0 & 0
\end{pmatrix}
\begin{pmatrix}
    k & l_3-\frac{1}{2} & l_4+\frac{1}{2} \\ n & -m_3-\frac{1}{2} & m_4+\frac{1}{2}
\end{pmatrix}\nn\\
&\quad &+i^{l_4-l_3}(-1)^{m_3-\frac{1}{2}}s_3s_4\sqrt{(l_3-m_3+1)(l_4-m_4+1)}
\begin{pmatrix}
    k & l_3+\frac{1}{2} & l_4+\frac{1}{2} \\ 0 & 0 & 0
\end{pmatrix}
\begin{pmatrix}
    k & l_3+\frac{1}{2} & l_4+\frac{1}{2} \\ n & -m_3+\frac{1}{2} & m_4-\frac{1}{2}
\end{pmatrix}\nn\\
&\quad &+i^{l_4-l_3}(-1)^{m_3+\frac{1}{2}}s_1s_4\sqrt{(l_3+m_3+1)(l_4+m_4+1)}
\begin{pmatrix}
    k & l_3+\frac{1}{2} & l_4+\frac{1}{2} \\ 0 & 0 & 0
\end{pmatrix}
\begin{pmatrix}
    k & l_3+\frac{1}{2} & l_4+\frac{1}{2} \\ n & -m_3-\frac{1}{2} & m_4+\frac{1}{2}
\end{pmatrix}.
\eeq
Here $\begin{pmatrix}
    l & l^\prime & l^{\prime\prime} \\ m & m^\prime & m^{\prime\prime}
\end{pmatrix}$ is the Wigner $3j$-symbol which is non-vanishing only when $m+m^\prime+m^{\prime\prime}=0$. Thus the conservation of angular momenta is manifest, i.e. $m_2-m_1=n=m_3-m_4$, and $(-1)^{n+m_1-\frac{1}{2}+m_3-\frac{1}{2}}=-(-1)^{m_2+m_3}$. The interaction Hamiltonian can be rearranged as
\bea
H_\mathrm{int}=
\sum_{\substack{s_1,s_2,s_3,s_4 \\ l_1,l_2,l_3,l_4 \\ m_1,m_2,m_3,m_4}}&\big(f^\dagger_{s_1,l_1,m_1}M f_{s_2,l_2,m_2}\big)\big(f^\dagger_{s_3,l_3,m_3}M f_{s_4,l_4,m_4}\big)s_2 s_3 V_{s_1,-s_2,-s_3,s_4;l_1,l_2,l_3,l_4;m_1,m_2,m_3,m_4}\\
&+\big(f^\dagger_{s_1,l_1,m_1}Q f_{s_2,l_2,m_2}\big)\big(f^\dagger_{s_3,l_3,m_3}Q f_{s_4,l_4,m_4}\big)V^\prime_{s_1,s_2,s_3,s_4;l_1,l_2,l_3,l_4;m_1,m_2,m_3,m_4},
\eea
where the interacting potential reads
\beq
&&V_{s_1,s_2,s_3,s_4;l_1,l_2,l_3,l_4;m_1,m_2,m_3,m_4}\nn\\
&=&\sum_{k,n}-\frac{4\pi V_k}{2k+1}\int\mathrm{d}\Omega_a~Y^*_{k,n}(\Omega_a)\Upsilon^\dagger_{s_1,l_1,m_1}(\Omega_a)\Upsilon_{s_2,l_2,m_2}(\Omega_a)\int\mathrm{d}\Omega_b Y_{k,n}(\Omega_b)\Upsilon^\dagger_{s_3,l_3,m_3}(\Omega_b)\Upsilon_{s_4,l_4,m_4}(\Omega_b)\nn\\
&=&\sum_k V_k\, i^{-l_1+l_2-l_3+l_4} (-1)^{m_2+m_3}\delta_{m_1-m_2+m_3-m_4,0}\times\nn\\
&\quad &\quad~\Bigg[\sqrt{(l_1+m_1)(l_2+m_2)}
\begin{pmatrix}
    k & l_1-\frac{1}{2} & l_2-\frac{1}{2} \\ 0 & 0 & 0
\end{pmatrix}
\begin{pmatrix}
    k & l_1-\frac{1}{2} & l_2-\frac{1}{2} \\ m_1-m_2 & -m_1+\frac{1}{2} & m_2-\frac{1}{2}
\end{pmatrix}\nn\\
&\quad &\qquad -\sqrt{(l_1-m_1)(l_2-m_2)}
\begin{pmatrix}
    k & l_1-\frac{1}{2} & l_2-\frac{1}{2} \\ 0 & 0 & 0
\end{pmatrix}
\begin{pmatrix}
    k & l_1-\frac{1}{2} & l_2-\frac{1}{2} \\ m_1-m_2 & -m_1-\frac{1}{2} & m_2+\frac{1}{2}
\end{pmatrix}\nn\\
&\quad &\qquad -is_1\sqrt{(l_1-m_1+1)(l_2+m_2)}
\begin{pmatrix}
    k & l_1+\frac{1}{2} & l_2-\frac{1}{2} \\ 0 & 0 & 0
\end{pmatrix}
\begin{pmatrix}
    k & l_1+\frac{1}{2} & l_2-\frac{1}{2} \\ m_1-m_2 & -m_1+\frac{1}{2} & m_2-\frac{1}{2}
\end{pmatrix}\nn\\
&\quad &\qquad -is_1\sqrt{(l_1+m_1+1)(l_2-m_2)}
\begin{pmatrix}
    k & l_1+\frac{1}{2} & l_2-\frac{1}{2} \\ 0 & 0 & 0
\end{pmatrix}
\begin{pmatrix}
    k & l_1+\frac{1}{2} & l_2-\frac{1}{2} \\ m_1-m_2 & -m_1-\frac{1}{2} & m_2+\frac{1}{2}
\end{pmatrix}\nn\\
&\quad &\qquad +is_2\sqrt{(l_1+m_1)(l_2-m_2+1)}
\begin{pmatrix}
    k & l_1-\frac{1}{2} & l_2+\frac{1}{2} \\ 0 & 0 & 0
\end{pmatrix}
\begin{pmatrix}
    k & l_1-\frac{1}{2} & l_2+\frac{1}{2} \\ m_1-m_2 & -m_1+\frac{1}{2} & m_2-\frac{1}{2}
\end{pmatrix}\nn\\
&\quad &\qquad +is_2\sqrt{(l_1-m_1)(l_2+m_2+1)}
\begin{pmatrix}
    k & l_1-\frac{1}{2} & l_2+\frac{1}{2} \\ 0 & 0 & 0
\end{pmatrix}
\begin{pmatrix}
    k & l_1-\frac{1}{2} & l_2+\frac{1}{2} \\ m_1-m_2 & -m_1-\frac{1}{2} & m_2+\frac{1}{2}
\end{pmatrix}\nn\\
&\quad &\qquad +s_1s_2\sqrt{(l_1-m_1+1)(l_2-m_2+1)}
\begin{pmatrix}
    k & l_1+\frac{1}{2} & l_2+\frac{1}{2} \\ 0 & 0 & 0
\end{pmatrix}
\begin{pmatrix}
    k & l_1+\frac{1}{2} & l_2+\frac{1}{2} \\ m_1-m_2 & -m_1+\frac{1}{2} & m_2-\frac{1}{2}
\end{pmatrix}\nn\\
&\quad &\qquad -s_1s_2\sqrt{(l_1+m_1+1)(l_2+m_2+1)}
\begin{pmatrix}
    k & l_1+\frac{1}{2} & l_2+\frac{1}{2} \\ 0 & 0 & 0
\end{pmatrix}
\begin{pmatrix}
    k & l_1+\frac{1}{2} & l_2+\frac{1}{2} \\ m_1-m_2 & -m_1-\frac{1}{2} & m_2+\frac{1}{2}
\end{pmatrix}\Bigg]\times\nn\\
&\quad &\quad~ \Bigg[ \sqrt{(l_3+m_3)(l_4+m_4)}
\begin{pmatrix}
    k & l_3-\frac{1}{2} & l_4-\frac{1}{2} \\ 0 & 0 & 0
\end{pmatrix}
\begin{pmatrix}
    k & l_3-\frac{1}{2} & l_4-\frac{1}{2} \\ m_3-m_4 & -m_3+\frac{1}{2} & m_4-\frac{1}{2}
\end{pmatrix}\nn\\
&\quad &\qquad -\sqrt{(l_3-m_3)(l_4-m_4)}
\begin{pmatrix}
    k & l_3-\frac{1}{2} & l_4-\frac{1}{2} \\ 0 & 0 & 0
\end{pmatrix}
\begin{pmatrix}
    k & l_3-\frac{1}{2} & l_4-\frac{1}{2} \\ m_3-m_4 & -m_3-\frac{1}{2} & m_4+\frac{1}{2}
\end{pmatrix}\nn\\
&\quad &\qquad -is_3\sqrt{(l_3-m_3+1)(l_4+m_4)}
\begin{pmatrix}
    k & l_3+\frac{1}{2} & l_4-\frac{1}{2} \\ 0 & 0 & 0
\end{pmatrix}
\begin{pmatrix}
    k & l_3+\frac{1}{2} & l_4-\frac{1}{2} \\ m_3-m_4 & -m_3+\frac{1}{2} & m_4-\frac{1}{2}
\end{pmatrix}\nn\\
&\quad &\qquad -is_3\sqrt{(l_3+m_3+1)(l_4-m_4)}
\begin{pmatrix}
    k & l_3+\frac{1}{2} & l_4-\frac{1}{2} \\ 0 & 0 & 0
\end{pmatrix}
\begin{pmatrix}
    k & l_3+\frac{1}{2} & l_4-\frac{1}{2} \\ m_3-m_4 & -m_3-\frac{1}{2} & m_4+\frac{1}{2}
\end{pmatrix}\nn\\
&\quad &\qquad +is_4\sqrt{(l_3+m_3)(l_4-m_4+1)}
\begin{pmatrix}
    k & l_3-\frac{1}{2} & l_4+\frac{1}{2} \\ 0 & 0 & 0
\end{pmatrix}
\begin{pmatrix}
    k & l_3-\frac{1}{2} & l_4+\frac{1}{2} \\ m_3-m_4 & -m_3+\frac{1}{2} & m_4-\frac{1}{2}
\end{pmatrix}\nn\\
&\quad &\qquad +is_4\sqrt{(l_3-m_3)(l_4+m_4+1)}
\begin{pmatrix}
    k & l_3-\frac{1}{2} & l_4+\frac{1}{2} \\ 0 & 0 & 0
\end{pmatrix}
\begin{pmatrix}
    k & l_3-\frac{1}{2} & l_4+\frac{1}{2} \\ m_3-m_4 & -m_3-\frac{1}{2} & m_4+\frac{1}{2}
\end{pmatrix}\nn\\
&\quad &\qquad +s_3s_4\sqrt{(l_3-m_3+1)(l_4-m_4+1)}
\begin{pmatrix}
    k & l_3+\frac{1}{2} & l_4+\frac{1}{2} \\ 0 & 0 & 0
\end{pmatrix}
\begin{pmatrix}
    k & l_3+\frac{1}{2} & l_4+\frac{1}{2} \\ m_3-m_4 & -m_3+\frac{1}{2} & m_4-\frac{1}{2}
\end{pmatrix}\nn\\
&\quad &\qquad -s_3s_4\sqrt{(l_3+m_3+1)(l_4+m_4+1)}
\begin{pmatrix}
    k & l_3+\frac{1}{2} & l_4+\frac{1}{2} \\ 0 & 0 & 0
\end{pmatrix}
\begin{pmatrix}
    k & l_3+\frac{1}{2} & l_4+\frac{1}{2} \\ m_3-m_4 & -m_3-\frac{1}{2} & m_4+\frac{1}{2}
\end{pmatrix}\Bigg],
\eeq
and similar for $V^\prime$ (where the only change is substituting $V_k$ to $V_k^\prime$). Note that such interacting potential satisfies
\beq
&&s_2 s_3 V_{s_1,-s_2,-s_3,s_4;l_1,l_2,l_3,l_4;m_1,m_2,m_3,m_4}=-s_2 s_4 V_{s_1,-s_2,s_3,-s_4;l_1,l_2,l_3,l_4;m_1,m_2,m_3,m_4}\nn\\
&=&s_1 s_4 V_{-s_1,s_2,s_3,-s_4;l_1,l_2,l_3,l_4;m_1,m_2,m_3,m_4}=-s_1 s_3 V_{-s_1,s_2,-s_3,s_4;l_1,l_2,l_3,l_4;m_1,m_2,m_3,m_4},\\
&&s_2 s_3 V_{s_1,-s_2,-s_3,s_4;l_1,l_2,l_3,l_4;m_1,m_2,m_3,m_4}=s_2 s_3 V^{*}_{s_4,-s_3,-s_2,s_1;l_4,l_3,l_2,l_1;m_4,m_3,m_2,m_1}\nn\\
&=&-s_2 s_4 V_{s_1,-s_2,s_3,-s_4;l_1,l_2,l_3,l_4;m_1,m_2,m_3,m_4}=-s_1 s_3 V^{*}_{s_4,-s_3,s_2,-s_1;l_4,l_3,l_2,l_1;m_4,m_3,m_2,m_1},
\eeq
as required by the convention we adopt and Hermiticity of the Hamiltonian, and is symmetric under particle-hole transformation. 

\subsection{Coordinate independent approach}

In this section, we solve for the eigenvalues and eigenfunctions of the Dirac operator on a sphere using symmetry considerations. The Dirac operator on any manifold is given by
\begin{equation}
    D = i \gamma^\mu \nabla_\mu,
\end{equation}
where $\nabla_\mu = \partial_\mu + \Omega_\mu$ is the covariant derivative for spinor fields, and $\Omega_\mu$ is the spin connection. On a sphere of radius $R$, the spin connection is constant due to the sphere's constant curvature. The Dirac operator simplifies to
\begin{equation}
    D = \frac{1}{R} \left( \boldsymbol{\sigma} \cdot \mathbf{L} + 1 \right),
\end{equation}
where $\boldsymbol{\sigma}$ are the Pauli matrices and $\mathbf{L}$ is the orbital angular momentum operator.

To find the eigenvalues of $D$, we use the relation $\mathbf{J} = \mathbf{L} + \frac{1}{2} \boldsymbol{\sigma}$ between the total angular momentum $\mathbf{J}$, the orbital angular momentum $\mathbf{L}$, and the spin $\boldsymbol{\sigma}/2$. Rewriting $\boldsymbol{\sigma} \cdot \mathbf{L}$ in terms of $\mathbf{J}^2$ and $\mathbf{L}^2$, we have
\begin{equation}
    \boldsymbol{\sigma} \cdot \mathbf{L} = \mathbf{J}^2 - \mathbf{L}^2 - \frac{3}{4}.
\end{equation}
Substituting this back into the Dirac operator, we get
\begin{equation}
    D = \frac{1}{R} \left( \mathbf{J}^2 - \mathbf{L}^2 + \frac{1}{4} \right).
\end{equation}

Let $\mathcal{Y}_{j m}^{\ell}$ denote the spinor spherical harmonics, which are eigenfunctions of both $\mathbf{J}^2$ and $\mathbf{L}^2$:
\begin{align}
    \mathbf{J}^2 \mathcal{Y}_{j m}^\ell &= j(j + 1) \mathcal{Y}_{j m}^\ell, \\
    \mathbf{L}^2 \mathcal{Y}_{j m}^\ell &= \ell(\ell + 1) \mathcal{Y}_{j m}^\ell,
\end{align}
where $j$ is the total angular momentum quantum number, $m$ is its projection, and $\ell$ is the orbital angular momentum quantum number. Using these relations, the action of the Dirac operator on $\mathcal{Y}_{j m}^\ell$ becomes
\begin{equation}
    D \mathcal{Y}_{j m}^\ell = \frac{1}{R} \left( j(j + 1) - \ell(\ell + 1) + \frac{1}{4} \right) \mathcal{Y}_{j m}^\ell.
\end{equation}
For two possible values of $\ell = j \pm \frac12$, we find the eigenvalues of the Dirac operator are
\begin{equation}
    D \mathcal{Y}_{j m}^\ell = \begin{cases}
    \dfrac{1}{R} \left( j + \dfrac{1}{2} \right) \mathcal{Y}_{j m}^\ell, & \ell = j - \dfrac{1}{2}, \\[2ex]
    -\dfrac{1}{R} \left( j + \dfrac{1}{2} \right) \mathcal{Y}_{j m}^\ell, & \ell = j + \dfrac{1}{2}.
    \end{cases}
\end{equation}
This result shows that the spinor spherical harmonics $\mathcal{Y}_{j m}^\ell$ are indeed eigenfunctions of the Dirac operator on the sphere, with eigenvalues determined by the total angular momentum quantum number $j$.

We aim to obtain the matrix elements of a generic two-body interaction operator \(\hat{V}\),
\begin{equation}
\left\langle j_1', m_1', \ell_1';\, j_2', m_2', \ell_2' \right| \hat{V} \left| j_1, m_1, \ell_1;\, j_2, m_2, \ell_2 \right\rangle = \int \mathcal{Y}_{j_1' m_1'}^{\ell_1' \dagger}(\mathbf{r}_1)\, \mathcal{Y}_{j_2' m_2'}^{\ell_2' \dagger}(\mathbf{r}_2)\, \hat{V}(\mathbf{r}_1, \mathbf{r}_2)\, \mathcal{Y}_{j_1 m_1}^{\ell_1}(\mathbf{r}_1)\, \mathcal{Y}_{j_2 m_2}^{\ell_2}(\mathbf{r}_2)\, d^3\mathbf{r}_1\, d^3\mathbf{r}_2
\end{equation}

Now we discuss two types of interactions. The first type factorizes into the spatial part and the spin part in the following way,
\begin{equation}
\hat{V}(\mathbf{r}_1, \mathbf{r}_2) = V\left( |\mathbf{r}_1 - \mathbf{r}_2| \right)\, \hat{O}^{(1)}_{m_{s_1}', m_{s_1}} \otimes \hat{O}^{(2)}_{m_{s_2}', m_{s_2}}
\end{equation}
Then we expand the potential \( V\left( |\mathbf{r}_1 - \mathbf{r}_2| \right) \) using spherical harmonics,
\begin{equation}
V\left( |\mathbf{r}_1 - \mathbf{r}_2| \right) = \sum_{k=0}^\infty \sum_{m=-k}^k \frac{V_k}{2k+1} Y_{k m}^*(\Omega_1)\, Y_{k m}(\Omega_2),
\end{equation}
where \( Y_{k m}(\Omega) \) are the spherical harmonics, and \( \Omega_i \) denotes the angular coordinates of \( \mathbf{r}_i \). This expansion allows us to perform the integrations over \( \mathbf{r}_1 \) and \( \mathbf{r}_2 \) independently.

Since \( \hat{O} \) acts only on the spinors, we decompose the spinor spherical harmonics into their orbital and spin parts,
\begin{equation}
\mathcal{Y}_{j m}^{\ell}(\theta, \phi) = \sum_{m_\ell, m_s} \left\langle \ell\, \tfrac{1}{2}\, m_\ell\, m_s \mid j\, m \right\rangle\, Y_{\ell m_\ell}(\theta, \phi)\, \chi_{m_s}
\end{equation}
where \( \left\langle \ell\, \tfrac{1}{2}\, m_\ell\, m_s \mid j\, m \right\rangle \) are the Clebsch-Gordan coefficients. We evaluate the spinor matrix elements as
\begin{equation}
\chi_{m_{s}'}^{\dagger}\, \hat{O}^{(i)}\, \chi_{m_{s}} = \sigma_a^{m_{s}', m_{s}}
\end{equation}
for \( \hat{O}^{(i)} = \sigma_a \).

Next, we handle the integration over the solid angles. The integral involving three spherical harmonics is given by
\begin{equation}
\int Y_{\ell' m_{\ell'} }^*(\Omega)\, Y_{\ell m_\ell}(\Omega)\, Y_{k m}^*(\Omega)\, d\Omega = (-1)^{m_{\ell'}+m} \sqrt{ \frac{ (2\ell' + 1)(2\ell + 1)(2k + 1) }{4\pi} }\, \begin{pmatrix}
\ell' & \ell & k \\
0 & 0 & 0
\end{pmatrix}\, \begin{pmatrix}
\ell' & \ell & k \\
-m_{\ell'} & m_\ell & -m
\end{pmatrix}
\end{equation}
We can express the Clebsch-Gordan coefficients using Wigner 3j symbols,
\begin{equation}
\left\langle \ell\, \tfrac{1}{2}\, m_\ell\, m_s \mid j\, m \right\rangle = (-1)^{\ell - m_\ell + \tfrac{1}{2} - m_s}\, \sqrt{2j + 1}\, \begin{pmatrix}
\ell & \tfrac{1}{2} & j \\
m_\ell & m_s & -m
\end{pmatrix}
\end{equation}

Putting everything together, we find
\begin{equation}
\begin{aligned}
& \left\langle j_1', m_1', \ell_1';\, j_2', m_2', \ell_2' \left| \hat{V} \right| j_1, m_1, \ell_1;\, j_2, m_2, \ell_2 \right\rangle =  \sum_{\{ m_s \}}\, \sum_{k=0}^\infty \,  \frac{V_k}{\left(2k + 1\right)^2}  \\
& \times \sqrt{ (2j_1' + 1)(2j_1 + 1) }\, \sqrt{ (2j_2' + 1)(2j_2 + 1) }\, \sqrt{ (2\ell_1' + 1)(2\ell_1 + 1)(2k + 1) }\, \sqrt{ (2\ell_2' + 1)(2\ell_2 + 1)(2k + 1) } \\
& \times (-1)^{ (\ell_1' + \ell_1 - m_1' - m_{s_1}) + (\ell_2' + \ell_2 - m_2 - m_{s_2}') } \\
& \times \begin{pmatrix}
\ell_1' & \tfrac{1}{2} & j_1' \\
m_1' - m_{s_1}' & m_{s_1'} & -m_1'
\end{pmatrix}\, \begin{pmatrix}
\ell_1 & \tfrac{1}{2} & j_1 \\
m_1 - m_{s_1} & m_{s_1} & -m_1
\end{pmatrix}\, \begin{pmatrix}
\ell_1' & \ell_1 & k \\
0 & 0 & 0
\end{pmatrix}\, \begin{pmatrix}
\ell_1' & \ell_1 & k \\
-m_1' + m_{s_1'} & m_1 - m_{s_1} & -m
\end{pmatrix} \\
& \times \begin{pmatrix}
\ell_2' & \tfrac{1}{2} & j_2' \\
m_2' - m_{s_2'} & m_{s_2'} & -m_2'
\end{pmatrix}\, \begin{pmatrix}
\ell_2 & \tfrac{1}{2} & j_2 \\
m_2 - m_{s_2} & m_{s_2} & -m_2
\end{pmatrix}\, \begin{pmatrix}
\ell_2' & \ell_2 & k \\
0 & 0 & 0
\end{pmatrix}\, \begin{pmatrix}
\ell_2' & \ell_2 & k \\
-m_2' + m_{s_2'} & m_2 - m_{s_2} & m
\end{pmatrix} \\
& \times \sigma_a^{m_{s_1}', m_{s_1}}\, \sigma_b^{m_{s_2}', m_{s_2}}\, \delta_{ -m_1' + m_1 - m_2' + m_2 + m_{s_1'} - m_{s_1} + m_{s_2'} - m_{s_2} }
\end{aligned}
\end{equation}
where \( m = (m_1 - m_{s_1}) - (m_1' - m_{s_1'}) \), and \( \hat{O}^{(1)} = \sigma_a \), \( \hat{O}^{(2)} = \sigma_b \). Here we have simplified the result using quantum number conservation from each 3j symbol.


The matrix elements of one-body potential $\hat{\mu}$ is much simpler and can be obtained using a similar approach.
\begin{equation}
    \left\langle j^{\prime}, m^{\prime}, \ell^{\prime}\right| \hat{V}\left|j, m, \ell\right\rangle=\int \mathcal{Y}_{j^{\prime} m^{\prime}}^{\ell^{\prime} \dagger}\left(\mathbf{r}\right) \hat{\mu}\left(\mathbf{r}\right) \mathcal{Y}_{j m}^{\ell}\left(\mathbf{r}\right) d^3 \mathbf{r}
\end{equation}
Suppose $\hat{\mu}$ factorizes into $\hat{\mu} = \mu(\mathbf{r}) \hat{O}$. The spin part still gives me $\chi_{m_{s_i}'}^{\dagger}\, \hat{O}\, \chi_{m_{s_i}} = \sigma_a^{m_{s_i}', m_{s_i}}$, while the spatial part requires a different decomposition,
\begin{equation}
    \mu(\Omega)=\sum_{k=0}^{\infty} \sum_{m=-k}^{k} \mu_{k m} Y_{k m}(\Omega)
\end{equation}
Then we can obtain the matrix elements as follows,
\begin{equation}
    \begin{aligned}
&\left\langle j', m', \ell' \left| \hat{\mu} \right| j, m, \ell \right\rangle = \sum_{k} \mu_{k,m' - m_{s}' - m + m_s}\, (-1)^{(\ell' + \ell + 1) - (m_{s}' + m)} \sqrt{ (2j' + 1)(2j + 1) } \sqrt{ \frac{ (2\ell' + 1)(2k + 1)(2\ell + 1) }{ 4\pi } } \\
&\times \begin{pmatrix}
\ell' & \tfrac{1}{2} & j' \\
m' - m_{s}' & m_{s}' & -m'
\end{pmatrix} \begin{pmatrix}
\ell & \tfrac{1}{2} & j \\
m - m_s & m_{s} & -m
\end{pmatrix} \begin{pmatrix}
\ell' & \ell & k \\
0 & 0 & 0
\end{pmatrix} \begin{pmatrix}
\ell' & \ell & k \\
-m' + m_{s}' & m - m_s & m' - m_{s}' - m + m_s
\end{pmatrix} \sigma_a^{m_{s}', m_{s}}
\end{aligned}
\end{equation}

One might also want to use another type of interactions, where the spin part rotates on the sphere,
\begin{equation}
\hat{V}(\mathbf{r}_1, \mathbf{r}_2) = V\left( |\mathbf{r}_1 - \mathbf{r}_2| \right)\, \hat{O}^{(1)}_{m_{s_1}', m_{s_1}}(\mathbf{r}_1) \otimes \hat{O}^{(2)}_{m_{s_2}', m_{s_2}}(\mathbf{r}_2)
\end{equation}
In particular we will consider $\hat{O}(\mathbf{r}) = \mathbf{n}(\Omega) \cdot \boldsymbol{\sigma}$ pointing in the normal direction of sphere. In this case, most of the calculation follows, but the spinor matrix elements become
\begin{equation}
\chi_{m_{s}'}^{\dagger}\, \hat{O}\, \chi_{m_{s}} = \left(\begin{array}{cc}
\cos \theta & \sin \theta e^{-i \phi} \\
\sin \theta e^{i \phi} & -\cos \theta
\end{array}\right)_{m_{s}', m_{s}} = \sqrt{\frac{4 \pi}{3}} \left(\begin{array}{cc}
Y_{1,0} & \sqrt{2} Y_{1,-1} \\
-\sqrt{2} Y_{1,1} & -Y_{1,0}
\end{array}\right)_{m_{s}', m_{s}}
\end{equation}
Now we need to include these factors in the integration of spherical harmonics,
\begin{equation}
\begin{aligned}
&\int Y_{\ell' m_{\ell'} }^*(\Omega)\, Y_{\ell m_\ell}(\Omega)\, Y_{k m}^*(\Omega) Y_{1q}(\Omega)\, d\Omega = (-1)^{m_{\ell}+m} 
\sum_{l} \frac{\sqrt{3\left(2 \ell'+1\right)\left(2 \ell+1\right)\left(2 k+1\right)}}{4 \pi}(2 l+1) \\ & \qquad \qquad \qquad \qquad \times \left(\begin{array}{ccc}
\ell' & \ell & l \\
0 & 0 & 0
\end{array}\right)\left(\begin{array}{ccc}
k & 1 & l \\
0 & 0 & 0
\end{array}\right)\left(\begin{array}{ccc}
\ell' & \ell & l \\
-m_{\ell}' & m_{\ell} & q - m
\end{array}\right)\left(\begin{array}{ccc}
k & 1 & l \\
-m & q & m - q
\end{array}\right) \delta_{m_{\ell}' - m_{\ell} + m - q}
\end{aligned}
\end{equation}
Now putting everything together we find
Putting everything together, we find
\begin{equation}
\begin{aligned}
& \left\langle j_1', m_1', \ell_1';\, j_2', m_2', \ell_2' \left| \hat{V} \right| j_1, m_1, \ell_1;\, j_2, m_2, \ell_2 \right\rangle = \sum_{\{ m_s \},l_1,l_2}\, \sum_{k=0}^\infty\, \frac{V_k}{\left(2k + 1\right)^2}  \\
& \times \sqrt{ (2j_1' + 1)(2j_1 + 1) }\, \sqrt{ (2j_2' + 1)(2j_2 + 1) }\, \sqrt{ (2\ell_1' + 1)(2\ell_1 + 1)(2k + 1) }\, \sqrt{ (2\ell_2' + 1)(2\ell_2 + 1)(2k + 1) } \\
& \times (2l_1+1) (2l_2+1) (-1)^{ (\ell_1' + \ell_1 + m_1 + m_{s_1}) + (\ell_2' + \ell_2 + m_2' + m_{s_2}) } \\
& \times \begin{pmatrix}
\ell_1' & \tfrac{1}{2} & j_1' \\
m_1' - m_{s_1'} & m_{s_1'} & -m_1'
\end{pmatrix}\, \begin{pmatrix}
\ell_1 & \tfrac{1}{2} & j_1 \\
m_1 - m_{s_1} & m_{s_1} & -m_1
\end{pmatrix} \, \begin{pmatrix}
\ell_2' & \tfrac{1}{2} & j_2' \\
m_2' - m_{s_2'} & m_{s_2'} & -m_2'
\end{pmatrix}\, \begin{pmatrix}
\ell_2 & \tfrac{1}{2} & j_2 \\
m_2 - m_{s_2} & m_{s_2} & -m_2
\end{pmatrix} \\
& \times \begin{pmatrix}
\ell_1' & \ell_1 & l_1 \\
0 & 0 & 0
\end{pmatrix}\, \begin{pmatrix}
k & 1 & l_1 \\
0 & 0 & 0
\end{pmatrix}\, \begin{pmatrix}
\ell_1' & \ell_1 & l_1 \\
-m_1' + m_{s_1}' & m_1 - m_{s_1} & m_{s_1} - m_{s_1}' -m
\end{pmatrix} \, \begin{pmatrix}
k & 1 & l_1 \\
-m & m_{s_1} - m_{s_1}'  & m + m_{s_1}' - m_{s_1}
\end{pmatrix} \\
& \times \begin{pmatrix}
\ell_2' & \ell_2 & l_2 \\
0 & 0 & 0
\end{pmatrix}\, \begin{pmatrix}
k & 1 & l_2 \\
0 & 0 & 0
\end{pmatrix}\, \begin{pmatrix}
\ell_2' & \ell_2 & l_2 \\
-m_2' + m_{s_2}' & m_2 - m_{s_2} & m_{s_2} - m_{s_2}'  + m
\end{pmatrix} \, \begin{pmatrix}
k & 1 & l_2 \\
m & m_{s_2} - m_{s_2}'  & - m + m_{s_2}' - m_{s_2}
\end{pmatrix} \\
& \times (-1)^{1-m_{s_1}'-m_{s_2}'} \sqrt{2}^{|m_{s_1}' - m_{s_1}| + |m_{s_2}' - m_{s_2}|}\delta_{-m_1' + m_1 - m_2' + m_2}
\end{aligned}
\end{equation}
where \( m = m_1 - m_1' = - m_2 + m_2' \). Here we have used $q_1 = m_{s_1} - m_{s_1}'$ and  $q_2 = m_{s_2} - m_{s_2}'$ from the spinor matrix elements.

\section{Comparison of conformal data using velocity normalization of $\sigma$ and $T_{\mu\nu}$}

In this section we provide comparison of conformal data under
choices of velocity normalization using $\sigma$ and $T_{\mu\nu}$.

\setlength{\tabcolsep}{4.5pt}
\renewcommand{\arraystretch}{1.3}

\begin{table}[htbp]
\centering
\begin{tabular}{c | c c | c | c c}
\hline\hline
\(\mathcal{O}\)
& \(\Delta^{(\sigma)}\)
& \(\Delta^{(T)}\)
& \(\Delta_B\)
& \(\mathrm{Err}^{(\sigma)}\)
& \(\mathrm{Err}^{(T)}\)
\\
\hline

\(\sigma\)
& 0.66 & 0.70 & 0.66 & -- & 6.0\%
\\

\(\partial_\mu\sigma\)
& 1.64 & 1.74 & 1.66 & 1.2\% & 4.7\%
\\

\(\partial^2\sigma\)
& 2.81 & 2.98 & 2.66 & 5.6\% & 12.0\%
\\

\(\partial_\mu\partial_\nu\sigma\)
& 2.75 & 2.92 & 2.66 & 3.4\% & 9.6\%
\\

\hline

\(\psi\)
& 1.05 & 1.11 & 1.07 & 1.9\% & 4.0\%
\\

\(\partial_\mu\psi\)
& 2.04 & 2.16 & 2.07 & 1.4\% & 4.5\%
\\

\hline

\(\sigma_T\)
& 2.61 & 2.77 & 2.45 & 6.5\% & 12.9\%
\\

\(\partial_\mu\sigma_T\)
& 3.77 & 4.00 & 3.45 & 9.3\% & 15.8\%
\\

\hline

\multirow{2}{*}{\(\epsilon\)}
& \multirow{2}{*}{2.35}
& \multirow{2}{*}{2.49}
& 1.74 & 35\% & 43.2\%
\\
& & & 2.14 & 9.8\% & 16.4\%
\\

\hline

\(J_\mu\)
& 2.35 & 2.49 & 2.00 & 17\% & 24.6\%
\\

\(T_{\mu\nu}\)
& 2.83 & 3.00 & 3.00 & 5.7\% & --
\\

\hline

\(\widetilde{J}_\mu\)
& 2.39 & 2.53 & / & / & /
\\

\(D_{\mu\nu}\)
& 3.07 & 3.25 & / & / & /
\\

\hline\hline
\end{tabular}
\caption{%
Comparison of scaling dimensions at \(j_{\max}=3/2\) under two
choices of velocity normalization. Here \(\Delta^{(\sigma)}\) is
obtained by fixing \(\Delta_\sigma=0.66\), while
\(\Delta^{(T)}\) is obtained by fixing
\(\Delta_T=3\).
}
\label{tab:normalization_j32}
\end{table}

\setlength{\tabcolsep}{4.5pt}
\renewcommand{\arraystretch}{1.3}

\begin{table}[htbp]
\centering
\begin{tabular}{c | c c | c | c c}
\hline\hline
\(\mathcal{O}\)
& \(\Delta^{(\sigma)}\)
& \(\Delta^{(T)}\)
& \(\Delta_B\)
& \(\mathrm{Err}^{(\sigma)}\)
& \(\mathrm{Err}^{(T)}\)
\\
\hline

\(\sigma\)
& 0.66 & 0.70 & 0.66 & -- & 6.8\%
\\

\(\partial_\mu\sigma\)
& 1.66 & 1.77 & 1.66 & 0.0\% & 6.8\%
\\

\(\partial^2\sigma\)
& 2.68 & 2.86 & 2.66 & 0.8\% & 7.6\%
\\

\(\partial_\mu\partial_\nu\sigma\)
& 2.71 & 2.89 & 2.66 & 2.3\% & 8.8\%
\\

\hline

\(\psi\)
& 1.07 & 1.14 & 1.07 & 0.0\% & 6.8\%
\\

\(\partial_\mu\psi\)
& 1.97 & 2.10 & 2.07 & 4.8\% & 1.6\%
\\

\(\partial_\mu\partial_\nu\psi\)
& 3.08 & 3.29 & 3.07 & 0.3\% & 7.1\%
\\

\hline

\(\sigma_T\)
& 2.56 & 2.73 & 2.45 & 4.5\% & 11.6\%
\\

\(\partial_\mu\sigma_T\)
& 3.61 & 3.85 & 3.45 & 4.6\% & 11.7\%
\\

\hline

\multirow{2}{*}{\(\epsilon\)}
& \multirow{2}{*}{2.21}
& \multirow{2}{*}{2.36}
& 1.74 & 27\% & 35.6\%
\\
& & & 2.14 & 3.3\% & 10.3\%
\\

\hline

\(J_\mu\)
& 2.31 & 2.47 & 2.00 & 15\% & 23.3\%
\\

\(T_{\mu\nu}\)
& 2.81 & 3.00 & 3.00 & 6.3\% & --
\\

\hline

\(\widetilde{J}_\mu\)
& 2.31 & 2.47 & / & / & /
\\

\(D_{\mu\nu}\)
& 3.03 & 3.23 & / & / & /
\\

\hline\hline
\end{tabular}
\caption{%
Comparison of scaling dimensions at \(j_{\max}=5/2\) under two
choices of velocity normalization. Here \(\Delta^{(\sigma)}\) is
obtained by fixing \(\Delta_\sigma=0.66\), while
\(\Delta^{(T)}\) is obtained by fixing \(\Delta_T=3\).
}
\label{tab:normalization_j52}
\end{table}

\section{Nearly conserved operator $D_{\mu\nu}$ in $q=2$ sector}

We note that, for the free fermion CFT with flavor of Dirac fermion $N_f=1$, \emph{all} operators with scaling dimension $\Delta\le 4$ in $q=2$ sector can be classified into two conformal multiplets, the singlet pairing $\sigma_T$ and a $p$-wave pairing. Other primary operators will have scaling dimension $\Delta\ge 5$. The $\sigma_T$ multiplet is
\beq
&&\Delta=2,~\ell=0:~\sigma_T\sim \psi^\mathbf{T}R\psi,\\
&&\Delta=3,~\ell=1:~\partial_\mu\sigma_T\sim \partial_\mu\psi^\mathbf{T}R\psi,\\
&&\Delta=4,~\ell=0:~\partial^2\sigma_T\sim \partial_\mu\psi^\mathbf{T}R\partial_\mu\psi,\\
&&\Delta=4,~\ell=2:~\partial_\mu\partial_\nu\sigma_T\sim \partial_\mu\partial_\nu\psi^\mathbf{T}R\psi+\partial_\mu\psi^\mathbf{T}R\partial_\nu\psi.
\eeq
Here due to fermion statistics, $\chi^\mathbf{T}R\psi=\psi^\mathbf{T}R\chi$ is always satisfied. The $p$-wave pairing multiplet is
\beq
&&\Delta=3,~\ell=2:~\psi^\mathbf{T}R(\partial_{\mu}\gamma_{\nu}+\partial_{\nu}\gamma_{\mu})\psi,\\
&&\Delta=4,~\ell=2:~\epsilon_{\rho\lambda\mu}\partial_\lambda [\psi^\mathbf{T}R(\partial_{\mu}\gamma_{\nu}+\partial_\nu\gamma_\mu)\psi],\\
&&\Delta=4,~\ell=3:~\partial_\lambda [\psi^\mathbf{T}R(\partial_{\mu}\gamma_{\nu}+\partial_{\nu}\gamma_{\mu})\psi].
\eeq
In particular, the $p$-wave pairing primary saturates the unitarity bound $\Delta=\ell+1$ for (2+1)D CFT, indicating it to be a conserved current. This is verified in its conformal tower without $\ell<2$ descendants. The $D_{\mu\nu}$ operator found in the GNY CFT also lives in the $q=2$ sector with angular momentum $\ell=2$, which is nearly conserved with scaling dimension very close to 3. Given these similarities, we conjecture $D_{\mu\nu}$ should be the $p$-wave pairing operator, $D_{\mu\nu}\sim \psi^\mathbf{T}R(\partial_{\mu}\gamma_{\nu}+\partial_{\nu}\gamma_{\mu})\psi$.

\vfill 

\end{document}